\documentclass[twocolumn,superscriptaddress,nofootinbib,amsmath,amssymb,amsfonts]{aastex631}

\usepackage{newtxtext,newtxmath}
\usepackage[T1]{fontenc}
\usepackage{ae,aecompl}
\usepackage{multirow}
\usepackage{rotating}
\usepackage{graphicx}	
\usepackage{amsmath}
\usepackage{url}
\usepackage{xcolor}
\usepackage{indentfirst}
\usepackage{ulem} 
\renewcommand{\sout}{\bgroup\markoverwith{\textcolor{black}{\rule[0.5ex]{2pt}{0.4pt}}}%
\ULon}

\begin{document}
\title{Neutron-quark stars: Discerning viable alternatives for the higher-density part of the equation of state of compact stars}
\author{Sudipta Hensh}
\email{sudiptahensh2009@gmail.com}
\affiliation{Department of Earth and Space Science, Graduate School of Science, The University of Osaka, 1-1 Machikaneyama-cho, Toyonaka, Osaka 560-0043, Japan}
\affiliation{Astrophysical Big Bang Laboratory (ABBL), RIKEN Pioneering Research Institute (PRI), 2-1 Hirosawa, Wako, Saitama 351-0198, Japan}

\author{Yong-Jia Huang}
\email{huangyj@pmo.ac.cn}
\affiliation{Key Laboratory of Dark Matter and Space Astronomy, Purple Mountain Observatory, Chinese Academy of Science, Nanjing, 210008, China}
\affiliation{RIKEN Center for Interdisciplinary Theoretical and Mathematical Sciences (iTHEMS), RIKEN, Wako 351-0198, Japan}

\author{Toru Kojo}
\email{toru.kojo.b1@tohoku.ac.jp}
\affiliation{Department of Physics, Tohoku University, Sendai, 980-8578, Japan}
\affiliation{KEK Theory Center, Institute of Particle and Nuclear Studies, High Energy Accelerator Research Organization, 1-1 Oho, Tsukuba, Ibaraki 305-0801, Japan}

\affiliation{Graduate Institute for Advanced Studies, SOKENDAI, 1-1 Oho, Tsukuba, Ibaraki 305-0801, Japan}
\author{Luca Baiotti}
\affiliation{International College and Graduate School of Science, The University of Osaka, 1-2 Machikaneyama-cho, Toyonaka, Osaka 560-0043, Japan}
\author{Kentaro Takami}
\affiliation{Kobe City College of Technology, 651-2194 Kobe, Japan}
\affiliation{RIKEN Center for Interdisciplinary Theoretical and Mathematical Sciences (iTHEMS), RIKEN, Wako 351-0198, Japan}
\author{Shigehiro Nagataki}
\affiliation{Astrophysical Big Bang Laboratory (ABBL), RIKEN Pioneering Research Institute (PRI), 2-1 Hirosawa, Wako, Saitama 351-0198, Japan}
\affiliation{RIKEN Center for Interdisciplinary Theoretical and Mathematical Sciences (iTHEMS), RIKEN, Wako 351-0198, Japan}
\affiliation{RIKEN-Berkeley Center, RIKEN iTHEMS, University of California, Berkeley, Berkeley, CA 94720, USA}
\affiliation{Astrophysical Big Bang Group (ABBG), Okinawa Institute of Science and Technology (OIST), 1919-1 Tancha, Onna-son, Kunigami-gun, Okinawa 904-0495, Japan}
\author{Hajime Sotani}
\affiliation{Department of Mathematics and Physics, Kochi University, Kochi, 780-8520, Japan}
\affiliation{RIKEN Center for Interdisciplinary Theoretical and Mathematical Sciences (iTHEMS), RIKEN, Wako 351-0198, Japan}

\date{\today} 

\begin{abstract}
We investigate binary neutron star (BNS) mergers using general-relativistic numerical simulations with hadronic and hybrid equations of state (EOSs), incorporating the latest observations and theoretical constraints. 
We address two viable scenarios for the transition to quark matter: a quark-hadron crossover (QHC) or a strong first-order phase transition (1PT). 
To distinguish between different models, we define neutron-quark stars (NQS) as configurations where quark effects emerge at masses below the lowest observed neutron-star mass. While traditional “hybrid stars” may be distinguished by purely hadronic configurations through mass-radius measurements, the mass-radius relations of NQSs resemble those of purely hadronic models, with no sharp boundary between hadrons and quarks. The name NQS effectively captures the absence of a phase boundary between hadrons and quarks in QHC scenarios.
Our results indicate that QHC models can be distinguished from hadronic ones if both the inspiral and post-merger gravitational waves (GWs) are observed. In particular, the dominant post-merger frequency ($f_2$) tends to be lower than in hadronic models with the same tidal deformability ($\Lambda$).
We also present the first general-relativistic simulations of BNS mergers where the stars already contain quark matter before merging. These involve a strong first-order phase transition (1PT) at 1.8 times nuclear saturation density, followed by a stiff quark EOS. 
%In this case, although ($\Lambda$, compactness) deviates from hadronic relations, ($\Lambda$, $f_2$) aligns closely with hadronic EOS predictions. 
Finally, we identify a robust linear correlation between the total GW energy emitted after the merger and the $f_2$ frequency. Remarkably, this relation holds regardless of the quark presence.


\end{abstract}

\section{Introduction}
Neutron stars (NSs), composed of cold, dense matter at a few times nuclear saturation density $(n_0 \simeq 0.16\,\text{fm}^{-3})$, 
are cosmic laboratories for exploring dense matter in quantum chromodynamics (QCD)~\citep{Baym_2018,Lattimer:2021emm}. 
Recent progress in observations has significantly advanced our understanding of NS macroscopic properties, determining some of their characteristics with notable precision~\citep{PhysRevLett.119.161101,2019PhRvX...9a1001A,2019ApJ...887L..21R,2019ApJ...887L..24M,2021ApJ...918L..27R,2021ApJ...918L..28M,Choudhury:2024xbk,Tang:2024jvs}.
Combined with nuclear constraints at $\sim n_0$, the uncertainty in radius has been narrowed to approximately 1 km for NSs with masses in the range 1.4-2.0 $M_{\odot}$ at 90\% confidence level~\citep{Legred:2021hdx,2023SciBu..68..913H}. These observations begin to allow discrimination between EOSs containing different microscopic degrees of freedom.

Joint information, including the NICER mass-radius inference of PSR J0030+0451, PSR J0740+6620 and PSR J0437-4715~\citep{2019ApJ...887L..21R,2019ApJ...887L..24M,2021ApJ...918L..27R,2021ApJ...918L..28M,Choudhury:2024xbk}, tidal deformability from the GW170817 event~\citep{2019PhRvX...9a1001A}, and \textit{ab-initio} calculations from nuclear physics, such as chiral effective field theory (ChEFT)~\citep{2019PhRvL.122d2501D} near $n_0$ and perturbative QCD~(pQCD) at ultra-high densities~\citep{PhysRevD.81.105021,2021PhRvL.127p2003G}, have provided statistical constraints on the EOSs and the associated sound speed in a model-agnostic manner. The square of the sound speed $c_s^2$ has been found to be nonmonotonic as a function of density around 2-4 $n_0$, featuring at least one peak, exceeding the conformal limit ($c_s^2 =c^2/3$)~\citep{2023SciBu..68..913H,Altiparmak:2022bke,OmanaKuttan:2022aml,Brandes:2023hma,Annala:2023cwx}, suggesting that the matter within massive NSs is not purely hadronic. 
Indeed, in observationally preferred EOSs, the sound speed around 2$n_0$ grows with density at a rate greater than predicted by typical nucleonic descriptions.
This result arises from statistical analyses that take into account the existence of an NS with 2.08 $ M_{\odot} $~\citep{2020NatAs...4...72C} (or $\sim$ 2.3$M_{\odot}$, from the optical lightcurve of the black widow pulsar~\citep{2022ApJ...934L..17R}, or mass function~\citep{Shao:2020bzt, PhysRevD.109.043052}), the stringent theoretical and experimental constraints around $n_0$, and the fact that the radii for a 1.4 and a 2.1 $ M_{\odot}$ NS are found to be similar~\citep{2023SciBu..68..913H, Rutherford:2024srk, Tang:2024jvs}. The EOS softening in massive NSs, instead, stems from the pressure constraint from pQCD. 
Therefore, conventional 1PTs occurring at an intermediate density, where the observed $c_s^2$ peak is located, are disfavored, as such transitions involve a jump in energy density and thus induce EOS softening instead. To reconcile rapid stiffening with the expected appearance of quark degrees of freedom, quark-hadron crossover (QHC) scenarios are proposed~\citep{Masuda:2012ed,Baym:2019iky,Kojo:2021wax}, which, unlike 1PT models, stiffen the EOS during the hadron-quark transition~\citep{Kojo:2021ugu,PhysRevLett.132.112701}.

However, it is difficult to gain information on the high-density part of the stellar EOS only from observation of NSs in equilibrium, because a direct measurement of the maximum mass for a non-rotating NS
is hard, and such a measurement may not be sensitive to the state of matter at the NS center~\citep{Christian:2023hez}. Theoretical limits on the EOS depend somewhat on the implementation of pQCD constraints, which remain debated~\citep{Somasundaram:2021clp,Komoltsev:2023zor}. This causes the masquerade\footnote{{\it Masquerade effect} refers to the theoretical possibility that hybrid stars with quark-matter cores can appear indistinguishable from conventional NSs when judged by their external mass–radius relations, thus “masquerading” as NSs in observational data.} effect~\citep{Alford:2004pf} when an exotic state is present near the center of massive NSs.

\begin{figure}
    \centering
    \includegraphics[width=0.95\linewidth]{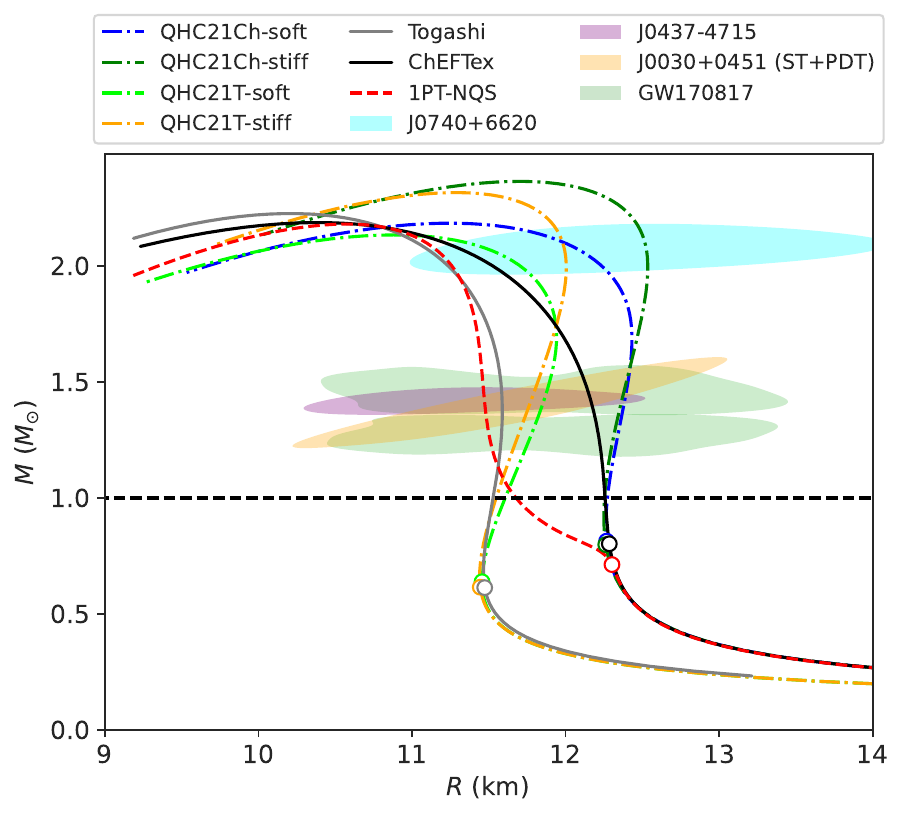}
    \includegraphics[width=0.95\linewidth]{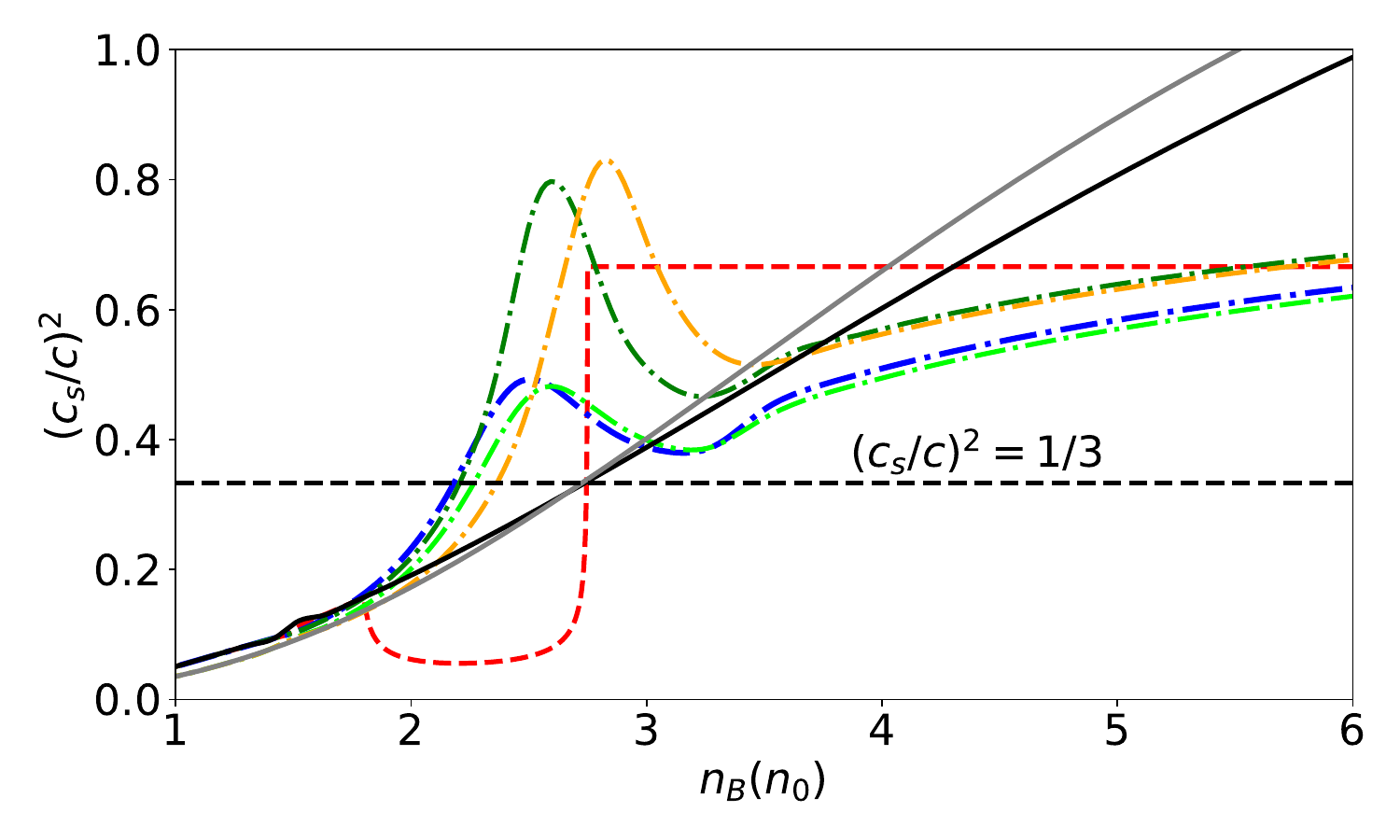}
    \caption{{\it Upper panel}: The $M$-$R$ relation for the NS and NQS models in this work, as well as the 68\% confidence interval from some observations. Open circles represent the theoretical NSs whose central density is $2 n_0$. {\it Lower panel}: The square of the sound speed as a function of baryon number density $n_B$ for the various EOSs. The dashed line represents the conformal limit $(c_s / c)^2 = 1/3$~\citep{PhysRevLett.114.031103}.}
    \label{fig:cs2}
\end{figure}

Another aspect of degeneracy in NS EOSs, as depicted in Fig.~\ref{fig:cs2}, is evidenced by the fact that EOSs with or without quark matter (and, thus, with different profiles of $c_s^2$ versus density) have a similar smooth structure in their mass-radius ($M$-$R$) curves above a solar mass. While exotic degrees of freedom certainly exist, their presence starting at low density results in property changes occurring only below the mass of the lightest NS, thus being masqueraded. To effectively describe such compact stars in which the hadron-quark transition significantly alters the original hadronic EOS, we propose the term "neutron-quark star" (NQS). In NQSs, the state of matter can be a mixture of neutrons and quarks, as in the case of a 1PT or a yet unknown state intermediate between nucleonic and quark matter, as proposed in QHC scenarios. The low-density transition is highlighted in the upper panel of Fig.~\ref{fig:cs2}, where open circles mark stars with central densities of $2n_0$. Crucially, within this density regime where the occurrence of a hadron-quark transition is not in conflict with constraints from nuclear calculations or experiments~\citep{Komoltsev:2024lcr,Tang:2025xib}, the masses for stars built with observationally favored EOSs remain sub-stellar (so below the lowest observed NS masses) at the time of the appearance of quark degrees of freedom. The name NQS indeed reflects the idea that, since the quarks appear at densities below the central density of the lightest NSs, quark degrees of freedom - whether partially or fully deconfined - are present in all compact stars.

In this letter, we study mergers of binaries~\citep{Bauswein2019,Weih:2019xvw,Most:2018eaw,HuangPRL2022,Raithel:2022efm,Fujimoto:2022xhv,Kedia:2022nns,Most:2022wgo,Zhu:2021xlu,Prakash:2023afe,Blacker:2024tet} composed of inspiralling NSs and NQSs. By incorporating the latest constraints from observations~\citep{2023SciBu..68..913H, Rutherford:2024srk, Tang:2024jvs}, we obtain more accurate information on the hadron-quark transition from the post-merger gravitational wave (GW) spectrum~\citep{Shibata:2003yj,Baiotti:2006wn,Xie:2020udh,Takami:2014tva,Takami:2014zpa,Rezzolla:2016nxn}. For the first time to the best of our knowledge, we present general-relativistic simulations of NQSs with a 1PT and the QHC EOS of~\cite{Kojo:2021wax}, in addition to NSs with purely nucleonic EOSs, for comparison. We study the hadron-quark-transition-dependent relation between $f_2$, the frequency of the main post-merger peak in the spectrum, and tidal deformability ($\Lambda$); and the hadron-quark-transition-independent relation between the total GW energy emitted in the post-merger phase and the average instantaneous frequency at the end of the simulations, when the remnant is quasi-stationary. These results are anticipated to resolve the degeneracy in static properties between NQSs and NSs when third-generation GW detectors — including Einstein Telescope~\citep{Prakash2021}, Cosmic Explorer~\citep{Reitze:2019iox}, and NEMO~\citep{Ackley:2020atn} — achieve their design sensitivities.

\section{Simulation setup and Equations of State}
A description of the codes we used and the setup of our simulations is in Appendix~\ref{Numerical Setup}. Here, we only mention that, conservatively, we assume the differences between the mean frequency of the $f_2$ peak ($f_{2,\textrm{mean}}$) at different resolutions to be our simulation numerical error on them. This is at most 50 Hz ({\it cf.} Table 1) and similar to the error due to the fitting procedure ({\it cf.} Fig.~\ref{fig:f2vslambda}).

As highlighted in Fig.~\ref{fig:cs2}, all the EOS models used here satisfy current multi-messenger constraints, including the maximum mass for a non-rotating NS,
and the tidal deformability and the radius
for a $1.4 M_{\odot}$ NS, according to the new NICER radius estimates for PSR J0437-4715 ($11.36^{+0.95}_{-0.63}$ km)~\citep{Choudhury:2024xbk} and PSR J0030+0451 ($11.71^{+0.88}_{-0.83}$ km in the ST+PDT model)~\citep{2024ApJ...961...62V}, which are more consistent with the findings from the GW170817 event~\citep{2019PhRvX...9a1001A} and slightly smaller than previous estimates~\citep{2019ApJ...887L..24M,2021ApJ...918L..28M}. The hadronic baseline\footnote{The {\it baseline} of an EOS containing quarks is the nucleonic EOS used for lower densities.} EOSs for our NQS models are either the Togashi EOS~\citep{Togashi:2017mjp} or an EOS based on the latest N3LO ChEFT results~\citep{Lonardoni2020, Drischler2021a, Drischler2021b}, which we name ChEFTex because we extended it to higher densities to be able to follow the merged object. Both of these are consistent with our current theoretical understanding of the EOS up to near $n_0$. Note that EOSs that differ significantly in the $c_s^2(n_B)$ relation (bottom panel of Fig.~\ref{fig:cs2}) are rather similar in the observable region of the $M$-$R$ plane.

The details of how the EOSs were constructed can be found in Appendix~\ref{Appendix-EOS-details}. Here, we briefly introduce the main features of our NQS models. In case a strong 1PT EOS has occurred, the radius of a high-mass NS is usually much smaller than that of a low-mass NS. In order for this to be compatible with current observations, such 1PT may occur either at the highest densities near the center of massive NSs or in a lower-density region and with a stiff quark-matter EOS~\citep{2023SciBu..68..913H}. 
Since we are interested in post-merger GWs, we consider here only the latter scenario; the former leads inevitably to prompt collapse to a black hole. Therefore, in our 1PT model, a quark-matter core is already present during the inspiral. The phase transition starts at $\sim 1.8 n_0$ and ends at $\sim 2.75 n_0$, at densities higher than which a stiff quark EOS $c_s^2 = 2/3~c^2$ is adopted. We call this EOS the 1PT-NQS EOS.

Another possible mechanism for the hadron-quark transition is QHC, whose main feature is the rapid stiffening and the associated sound-speed peak occurring in the region between the nuclear- and quark-matter domains (see Fig~\ref{fig:cs2}). 
The rapid stiffening leads to the radius of a $2.0 M_\odot$ NS being similar to or even larger than that of a $1.4M_{\odot}$ NS.

We consider two of the QHC21 models obtained with different parameters in~\cite{Kojo:2021wax}: the A model and the D model, which are, respectively, the softest and the stiffest in the sound-speed peak region. For each of them, we consider models with both alternatives for the baseline EOS: the Togashi EOS and the ChEFTex EOS. 
For these four models, we use the names QHC21Ch-soft, QHC21Ch-stiff, QHC21T-soft, and QHC21T-stiff, respectively.

We study equal-mass BNS systems with three total masses: {\it low} ($ 2.5 M_\odot$ for all models), {\it intermediate} ($ 2.75 M_\odot$ for all models), and {\it high} ($2.86 - 2.96 M_\odot$). 
The choice of the masses in the high range was made so that for each EOS, the collapse occurs late enough to accurately compute the post-merger frequency $f_2$. This was possible only for the QHC models in that mass range.

To avoid model-dependent phenomena in finite-temperature EOSs, we mimic thermal effects through an ideal-gas approximation with a fixed thermal index $\Gamma=2$, as in our previous work~\citep{HuangPRL2022} and as commonly done in the literature. 
In 1PT scenarios (as studied, {\it e.g.}, in~\cite{Blacker:2023afl,Blacker:2023opp}), the finite-temperature effects lower the onset density of the transition to quark matter (see, {\it e.g.},~\cite{Constantinou:2025wxj} and also the discussion in Appendix~\ref{Appendix-EOS-details}), potentially leading to a significantly higher $f_2$ than that of the baseline EOS~\citep{Fields:2023bhs}. However, how much the onset density is affected by finite-temperature effects is beyond the reach of observations of NSs in equilibrium. 
In this work, by adopting a fixed onset density through an ideal-gas thermal treatment, the simulation results reflect the properties of cold EOSs as constrained by current observations, in particular, providing a conservative estimate for the difference in $f_2$ in NSs and NQSs. We expect that finite-temperature effects  yield larger post-merger frequency differences.

\begin{figure}
    \centering\includegraphics[width=0.98\linewidth]{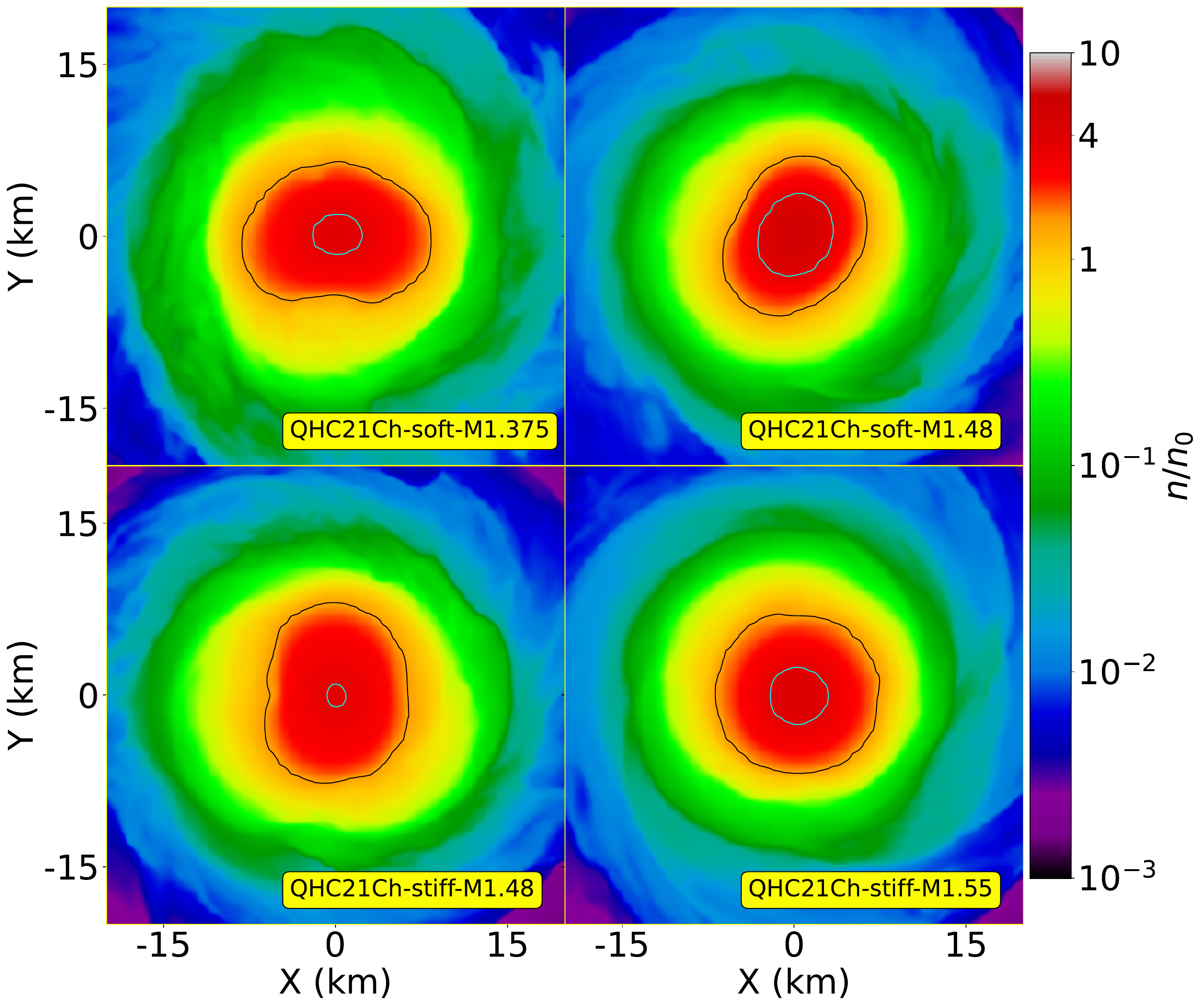}
    \caption{Isodensity contours on the (x,y) (equatorial) plane for the QHC21Ch models with different masses, as shown in each panel, at time $t = 10$ ms after the merger. 
    The black contour lines show the number density when the crossover starts. The cyan contour lines show the number density above which only deconfined quark matter is present. 
    }
    \label{fig:QHC21dens}
\end{figure}

\section{Results and discussion}
\subsection{Qualitative presentation and interpretation of the simulation results}

\begin{table*}
  \centering
\caption{Characteristics of the simulated BNS merger systems: The table presents various properties of the BNS merger systems studied in this work. 
The listed quantities, ordered from left to right, are: gravitational mass ($M_\textrm{star}$) of one isolated star, baryon mass ($M_\textrm{b star}$) of one star, radius ($R$) of one isolated non-rotating star, Arnowitt-Deser-Misner (ADM) mass of the BNS system~($M_\textrm{ADM}$) at the initial time, total angular momentum ($J$) at the initial time, orbital frequency ($f_\textrm{orb}$) at the initial time, dimensionless tidal deformability $\Lambda$, contact frequency ($f_\textrm{cont} = C^{3/2}/2\pi M$, where $C$ is compactness, namely the ratio of gravitational mass to the radius), post-merger peak frequency ($f_2$) with its 68\% confidence interval, and the mean frequency of the $f_2$ peak ($f_{2,\textrm{mean}}$). 
The units corresponding to these quantities are specified in the parentheses adjacent to each.
Simulations data refer to our medium resolution ($\Delta x \approx 231 \, \text{m}$) unless indicated in parentheses.}

\vspace{0.35cm}
    \scriptsize
    \setlength{\tabcolsep}{3pt}
    \begin{tabular*}{\textwidth}{@{\extracolsep{\fill}} l c c c c c c c c c c c c@{}}
    \hline
    \hline
    Model & \(M_*\) (\(M_\odot\)) & \(M_b\) (\(M_\odot\)) & \(R\) (km) & \(M_{\text{ADM}}\) (\(M_\odot\)) & \(J\) (\(M_\odot^2\)) & \(f_{\text{orb}}\) (Hz) & \(\Lambda\) & \(f_{\text{cont}}\) (Hz)& \(f_2\) (Hz)& \(f_{2,\text{mean}}\) (Hz)\\
    \hline
    QHC21Ch-soft-M1.25 & 1.25 & 1.36 & 12.32 & 2.48 & 6.4 & 273.03 & 854 & 1499.04 & \(2439^{+4}_{-4}\) & \(2849^{+9}_{-9}\) \\[0.5em]
    QHC21Ch-soft-M1.375 & 1.375 & 1.51 & 12.36 & 2.72 & 7.5 & 283.47 & 502 & 1564.59 & \(2619^{+4}_{-4}\) & \(3071^{+16}_{-15}\) \\[0.5em]
    QHC21Ch-soft-M1.375 (\(\Delta x \approx 185\, \text{m}\)) & & & & & & & & & \(2668^{+6}_{-5}\) & \(2955^{+8}_{-7}\)\\[0.5em]
    QHC21Ch-soft-M1.375 (\(\Delta x \approx 288\, \text{m}\)) & & & & & & & & & \(2629^{+9}_{-9}\) & \(3053^{+16}_{-15}\)\\[0.5em]
    QHC21Ch-soft-M1.48 & 1.48 & 1.64 & 12.39 & 2.92 & 8.48 & 291.69 & 329 & 1617.33 & \(2932^{+7}_{-7}\) & \(3165^{+13}_{-12}\) \\[0.5em]
    QHC21Ch-soft-M1.48 (\(\Delta x \approx 185\, \text{m}\)) & & & & & & & & & \(2881^{+21}_{-15}\) & \(3196^{+22}_{-25}\)\\[0.5em]
    QHC21Ch-stiff-M1.25 & 1.25 & 1.36 & 12.29 & 2.48 & 6.4 & 273.04 & 841 & 1504.54 & \(2511^{+28}_{-26}\) & \(2839^{+32}_{-20}\) \\[0.5em]
    QHC21Ch-stiff-M1.375 & 1.375 & 1.51 & 12.33 & 2.72 & 7.5 & 283.44 & 495 & 1570.3 & \(2634^{+33}_{-22}\) & \(2975^{+18}_{-21}\) \\[0.5em]
    QHC21Ch-stiff-M1.48 & 1.48 & 1.64 & 12.37 & 2.92 & 8.48 & 291.77 & 326 & 1623.23 & \(2688^{+8}_{-7}\) & \(3196^{+23}_{-21}\) \\[0.5em]
    QHC21Ch-stiff-M1.48 (\(\Delta x \approx 185\, \text{m}\)) & & & & & & & & & \(2754^{+6}_{-5}\) & \(3155^{+19}_{-18}\)\\[0.5em]
    QHC21Ch-stiff-M1.55 & 1.55 & 1.73 & 12.4 & 3.1 & 9.16 & 296.81 & 251 & 1653.14 & \(2818^{+13}_{-12}\) & \(3188^{+26}_{-24}\)\\[0.5em]
    QHC21T-soft-M1.25 & 1.25 & 1.37 & 11.77 & 2.48 & 6.4 & 273.02 & 693 & 1605.34 & \(2629^{+8}_{-7}\) & \(3008^{+10}_{-11}\)\\[0.5em]
    QHC21T-soft-M1.375 & 1.375 & 1.52 & 11.84 & 2.72 & 7.49 & 283.47 & 411 & 1668.78 & \(2816^{+12}_{-9}\) & \(3222^{+16}_{-19}\) \\[0.5em]
    QHC21T-soft-M1.43 & 1.43 & 1.59 & 11.87 & 2.82 & 8.00 & 287.79 & 329 & 1695.38 & \(2893^{+5}_{-6}\) & \(3257^{+20}_{-17}\)\\[0.5em]
    QHC21T-soft-M1.43 (\(\Delta x \approx 185\, \text{m}\)) & & & & & & & & & \(2923^{+280}_{-38}\) & \(3246^{+54}_{-58}\)\\[0.5em]
    QHC21T-stiff-M1.25 & 1.25 & 1.37 & 11.68 & 2.48 & 6.39 & 272.97 & 657 & 1623.93 & \(2600^{+13}_{-12}\) & \(3090^{+18}_{-18}\) \\[0.5em]
    QHC21T-stiff-M1.375 & 1.375 & 1.52 & 11.75 & 2.72 & 7.5 & 283.52 & 389 & 1687.99 & \(2784^{+13}_{-11}\) & \(3167^{+14}_{-15}\) \\[0.5em]
    QHC21T-stiff-M1.43 & 1.43 & 1.59 & 11.78 & 2.82 & 8.00 & 287.73 & 313 & 1714.85 & \(2962^{+13}_{-11}\) & \(3351^{+18}_{-17}\) \\[0.5em]
    QHC21T-stiff-M1.43 (\(\Delta x \approx 185\, \text{m}\)) & & & & & & & & & \(2750^{+11}_{-10}\) & \(3398^{+34}_{-33}\)\\[0.5em]
    Togashi-M1.25 & 1.25 & 1.37 & 11.57 & 2.47 & 6.39 & 273.00 & 587 & 1647.29 & \(2799^{+11}_{-10}\) & \(3170^{+15}_{-15}\) \\[0.5em]
    Togashi-M1.375 & 1.375 & 1.52 & 11.58 & 2.72 & 7.49 & 283.39 & 329 & 1898.25 & \(2976^{+8}_{-7}\) & \(3266^{+19}_{-17}\) \\[0.5em]
    ChEFTex-M1.25 & 1.25 & 1.36 & 12.21 & 2.48 & 6.39 & 272.93 & 784 & 1519.35 & \(2618^{+7}_{-8}\) & \(2932^{+11}_{-12}\) \\[0.5em]
    ChEFTex-M1.375 & 1.375 & 1.51 & 12.17 & 2.72 & 7.49 & 283.45 & 437 & 1526.84 & \(2940^{+5}_{-4}\) & \(3183^{+7}_{-7}\) \\[0.5em]
    1PT-NQS-M1.25 & 1.25 & 1.36 & 11.50 & 2.48 & 6.40 & 273.11 & 489 & 1662.20 & \(2896^{+5}_{-4}\) & \(3323^{+12}_{-13}\) \\[0.5em]
    1PT-NQS-M1.375 & 1.375 & 1.52 & 11.47 & 2.72 & 7.49 & 283.10 & 277 & 1750.18 & \(3051^{+21}_{-13}\) & \(3483^{+36}_{-42}\) \\[0.5em]
    1PT-NQS-M1.375 (\(\Delta x \approx 185\, \text{m}\)) & & & & & & & & & \(3072^{+7}_{-9}\) & \(3542^{+14}_{-14}\) \\[0.5em]
    1PT-NQS-M1.375 (\(\Delta x \approx 288\, \text{m}\)) & & & & & & & & & \(3038^{+7}_{-8}\) & \(3647^{+56}_{-54}\) \\[0.5em]
    \hline
    \hline
  \end{tabular*}
  \label{tab:allmodels}
\end{table*}

We first comment qualitatively on how observation of post-merger GWs may break the degeneracy between QHC models. In Fig.~\ref{fig:QHC21dens}, we show 2D isodensity contours 10 ms after the merger for two high-mass simulations with the QHC21Ch-stiff and QHC21Ch-soft EOSs, which feature the highest and lowest peak in $c_s^2$ among QHC21Ch models. The pure-hadron phase, crossover regime, and pure-quark phase of the QHC EOS are separated by the black and cyan contours. The different QHC21Ch models (here -stiff and -soft) mainly differ in the peak height of $c_s^2$, while their $M$-$R$ curves below $\sim$1.6$M_{\odot}$ are identical. 

We define {\it no-$f_2$ mass threshold} as the mass threshold above which the merger does not produce measurable $f_2$-mode GWs because its lifetime is too short. See Appendix~\ref{no-f2 mass threshold} for details and further comments.
The models in Fig.~\ref{fig:QHC21dens} are very near the no-$f_2$ mass threshold. The lifetimes for QHC21Ch-stiff-M1.48 and QHC21Ch-stiff-M1.55 are $> 25$ ms and $\sim 20$ ms, respectively, but simulations with stars of mass $1.60 M_{\odot}$ ($0.05 M_{\odot}$ higher than QHC21Ch-stiff-M1.55) result in a lifetime of less than 2 ms and no $f_2$-mode GW emission. We say this to highlight that the following considerations are almost independent of the considered masses.

Fig.~\ref{fig:QHC21dens} shows that for QHC21Ch-stiff, even for masses near the threshold, most of the star is in the crossover density region, rather than in the pure-quark region, implying that the details of the quark EOS have a smaller effect on the dynamics and value of $f_2$. For the QHC21Ch-soft EOSs, instead, the pure-quark EOS might have a predominant effect even if the mass is lower than that of the QHC21Ch-stiff models in the figure. 

One could think that for high enough masses, the soft quark EOS could lead to an $f_2$ higher than that expected from the inspiral, but from our results, stiff QHC models lead to universally lower post-merger GW frequency. 
The reason is that, given the expectation that the quark EOS is soft, models with a mass higher than those shown in Fig.~\ref{fig:QHC21dens} would collapse without $f_2$-mode GW emission and, consequently, the characteristic properties of a soft quark EOS would not manifest themselves in measurable post-merger GW signals. 
Therefore, for stiff QHC models, it is the crossover part of the EOS that has a dominant effect on dynamics and GW emission for all masses. 

From an observational perspective, this provides strong hints: an overly high $f_2$ frequency could strongly disfavor QHC EOSs with a prominent $c_s^2$ peak. In other words, we do not expect to observe an $f_2$ frequency higher than that expected from the inspiral if the EOS is a stiff QHC one, no matter how high the mass of the system is.
Also, observing an $f_2$ frequency much lower than expected from the inspiral could rule out the 1PT EOS (see below).

\subsection{EOS-dependent relation for $f_2$-$\Lambda$}
To connect the remnant properties to observation, we compute the spectrum of the fundamental and dominant harmonic mode ($l = m = 2$) of the GW strain up to $18$ ms after the merger for all models except for QHC21Ch-soft with mass $2.96 M_{\odot}$, where the collapse to black hole occurs around $11$ ms after the merger. 

The values of $f_2$ are estimated by fitting the spectrum through the Markov Chain Monte Carlo (MCMC) method described in the Supplemental Material of~\cite{HuangPRL2022}, in which the GW contribution from the inspiral is filtered out by approximating it with a power law plus exponential decay (starting at the contact frequency $f_{\rm cont} = C^{3/2}/(2\pi M)$, where $C$ represents stellar compactness~\citep{Damour-contact-frequency}; see Appendix~\ref{no-f2 mass threshold} for more details). The main peak is fitted by a skewed Gaussian because both amplitude and frequency evolve in time.
This MCMC framework enables statistically robust uncertainty quantification for the characteristic frequencies, and in this work, the fitting includes the $f_1$, $f_2$, and $f_3$ modes~\citep{Takami:2014zpa}. For a more direct comparison with other works, we also estimate the mean of the $f_2$ frequency, $f_{2, \text{mean}}$; the uncertainty is estimated by using the analytical expression of mean in a skewed Gaussian distribution $f_{2, \text{mean}} = f_{2} + \sigma \mu_z \sqrt{2/\pi}$, where $\sigma$ is the standard deviation, $\mu_z \equiv \alpha / \sqrt{1 + \alpha^2}$ is the often-used formula for the shift of the mode, and $\alpha$ is the model parameter that controls the direction of skewness. The uncertainties of the mean are taken to be the 68\% confidence interval of the distribution.

\begin{figure}[bh]
    \centering
        \includegraphics[width=0.95\linewidth]{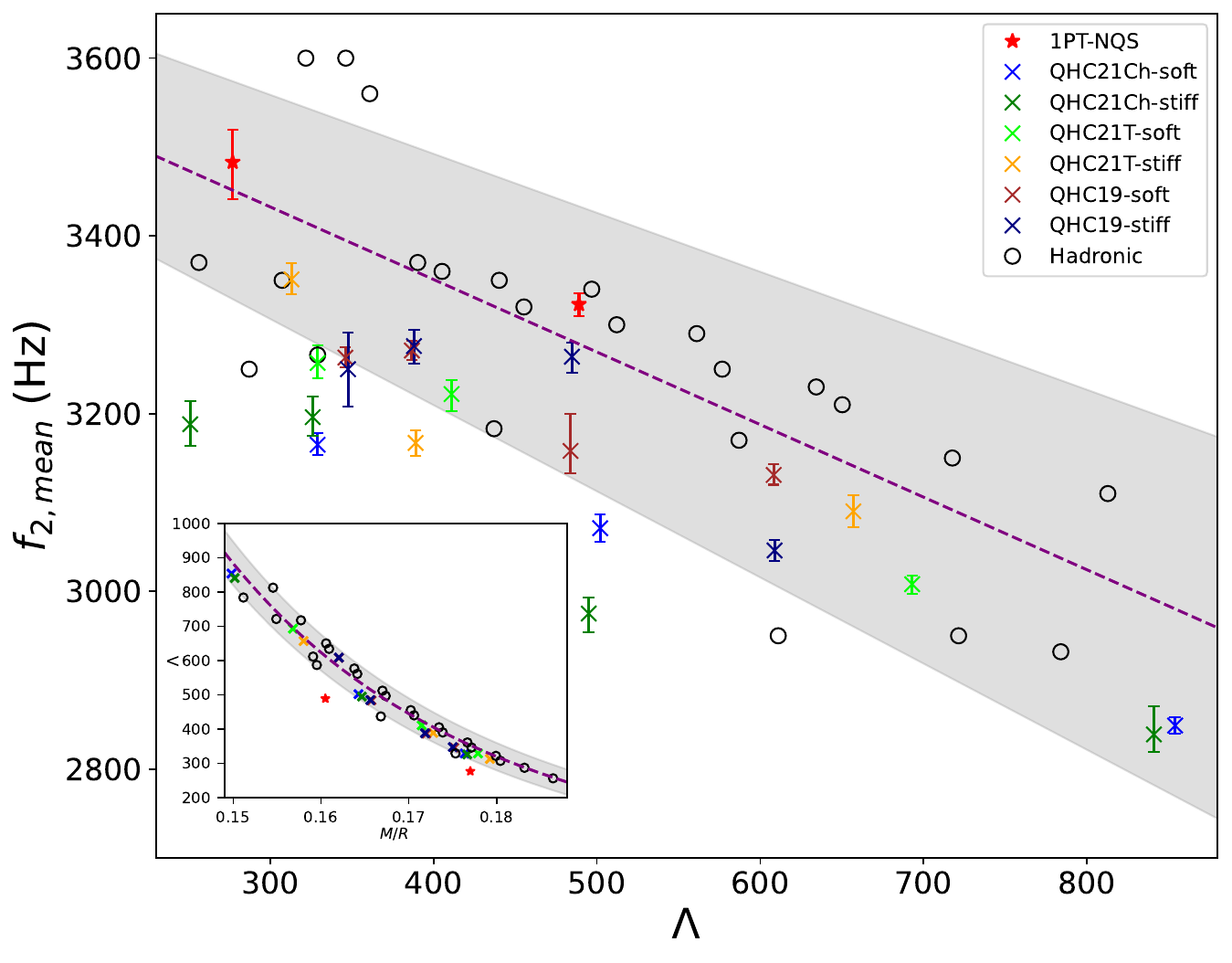}
        \caption{Scatter plot of $f_{2,\textrm{mean}}$ and 
        dimensionless tidal deformability.
        Hadronic EOSs include our Togashi and ChEFTex EOSs, and the SLy, APR4, and LS220 EOSs, from~\cite{Rezzolla:2016nxn}. Data for the QHC19 models are from~\cite{HuangPRL2022}.
        The dashed line shows the best-fit curve considering the hadronic models only. The shaded area shows the maximum residual of the fitting. The error bars indicate the fitting error in $f_{2,\text{mean}}$ for the NQS models.
        {\it Inset:} Scatter plot of the relation between the dimensionless tidal deformability and compactness for various EOSs.
        The dashed curve is the best fit considering the fitting relation, $\Lambda = \gamma + \frac{\delta}{(M/R)^5} \,$, where the parameters from the fit are $\gamma = -59.23 \pm 24.29$ and $\delta = 0.071 \pm 0.003$. The shaded area shows the maximum residual of fitting hadronic models.}
        \label{fig:f2vslambda}
\end{figure}

Although mass-radius curves of NSs and NQSs could be similar (see Fig.~\ref{fig:cs2}), joint information from before and after the merger can help discriminate between them. 
In Fig. \ref{fig:f2vslambda}, we plot the tidal deformability ($\Lambda$) versus $f_{2, \text{mean}}$ for both hadronic and NQS EOSs. For the former, we show the results for the ChEFTex EOS, the Togashi EOS, and the three EOSs, SLy~\citep{SLy}, APR4~\citep{APR}, and LS220~\citep{LS220}, studied in~\cite{Rezzolla:2016nxn} that are still compatible with current observational constraints~\citep{2023SciBu..68..913H}. We consider the linear fitting $f_{2, \text{mean}} = \alpha + \beta \, \Lambda$, where $\alpha$, and $\beta$ are parameters determined by fitting only the hadronic models. The best fits with their standard error are $\alpha = 3678.43 \pm 83.97$ Hz and $\beta = -0.83 \pm 0.16$ Hz, respectively. 
Note that the residual of the fitting of hadronic models will become smaller with further multi-messenger observations of NSs, since some of these models will be excluded by improved determination of the $M$-$R$ curve.

Inspecting the $f_{2, \text{mean}}$-$\Lambda$ relation, we notice that the QHC models show generally lower $f_{2, \text{mean}}$ than the best-fit curve for hadronic models. The QHC models with the strongest stiffening lie outside the lower bound of the one-standard-deviation error region 
by more than 200 Hz. Note that the numerical uncertainty in $f_{2, \text{mean}}$ due to finite resolution in our simulations is estimated to be at most 50 Hz~(this occurs for the QHC21Ch-soft-M1.375 model, with the highest resolution having a lower frequency; see Table 1). Also note that the uncertainty on tidal deformability is not important in this discussion, because it is much smaller. In fact, if the post-merger GWs are strong enough to be detected, the inspiral part of the same event will have an extremely high signal-to-noise ratio~(SNR), because the inspiral signal lies in the frequency band where detectors are most sensitive and accumulates over many cycles \citep{Flanagan1998PhRvD..57.4535F,Torres-Rivas2019PhRvD..99d4014T}.

The $f_{2, \text{mean}}$ for the 1PT-NQS models are about 300 Hz higher than their hadronic baseline (see Table 1), but the 1PT-NQS models are still compatible with the best-fit curve for the hadronic models. This is mainly because the stiffness of the 1PT-NQS models is similar before and after the merger, since in both phases, the stellar core is already in the density region where
$c_s = 2/3~c^2$. 
We emphasize again that ours is a conservative estimation, where a temperature-dependent phase boundary and more complex quark EOSs have not been considered.

Although the early presence of quark matter in the 1PT-NQS model may not be observable from the $\Lambda - f_2$ relation, the mass distribution for NQSs is undoubtedly different from that of NSs based on a nucleonic description. 
In the inset of Fig.~\ref{fig:f2vslambda}, we show the relation between $\Lambda$ and the compactness of a star with gravitational mass equal to that of one of the stars in the binary when isolated. 
We note that the 1PT-NQS EOS is an outlier in this relation (while the QHC EOSs lie near the best-fit curve). This means that, for example, the inferred radius from a $\Lambda$ measurement performed from GW detectors would be underestimated for the 1PT-NQS model. 
A measured $\Lambda$ that corresponds to a radius (measured, {\it e.g.}, by NICER) larger than that expected from the $\Lambda-$compactness relation would be in support of the 1PT-NQS scenario.

\begin{figure}
    \centering

\includegraphics[width=0.9\linewidth]{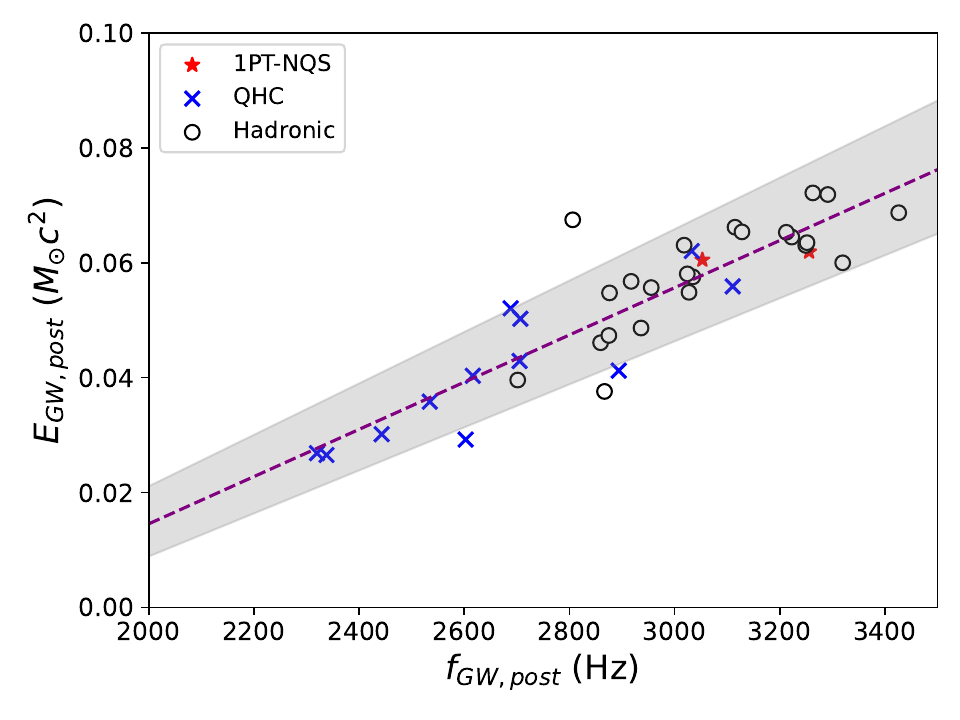}

    \caption{Scatter plot of GW energy released in the post-merger $E_{\text{GW,post}}$ and instantaneous frequency $f_{\text{GW,post}}$. $E_{\text{GW,post}}$ is calculated as the difference between the total energy emitted in GWs at the end of the simulation (when further emission is negligible, see Fig.~\ref{fig:angmom-energy-unnorm}) and that emitted until the time of the merger, from the $l=m=2$ mode (emission in other modes is negligible). $f_{\text{GW,post}}$ is the average GW frequency in an interval when the emission has become negligible, in practice, the last 5 ms of the simulation.
    The dashed line shows the best-fit curve.
    }
    \label{fig:E_fins}
\end{figure}

\subsection{Universal relation between GW energy emitted after the merger and post-merger frequency}
Another quantity of interest is the energy emitted in post-merger GWs and its relation to the nature of hadron-quark transitions.
Energy emission in post-merger GWs may be related to electromagnetic signals from binary NS/NQS mergers, and the latter may provide useful information for studying the central engine of $\gamma$-ray bursts~\citep{Rezzolla:2010fd,Fan:2013cra}. 
Since it is difficult to analytically describe the non-linear oscillations occurring after the merger, using our simulation results, we propose a method to estimate the energy emitted in GWs in the
post-merger phase, based only on a measurement of $f_2$. 
In fact, in actual GW observations, only $f_2$, and not the full waveform, may be measured with sufficient accuracy. Right after the merger, the GW frequency changes significantly because of the presence of multiple modes~\citep{Stergioulas2011b, Takami:2014zpa}, all of which are damped in a few ms, except for the $f_2$ mode.
GW energy is mainly released in the first $\approx 10$ ms after the merger, as after that time the remnant settles down to a less deformed configuration. Therefore, regardless of the remnant's lifetime (as long as it is not too short), the GW energy released in the first $\approx 20$ ms is a good estimate of the total released energy (see Fig.\ref{fig:angmom-energy-unnorm} in Appendix~\ref{Appendix-angmom-energy--details} for more details).
That is what we show in Fig.~\ref{fig:E_fins} for hadronic and NQS models: the scatter plot and the fitting of the energy released in GWs after the merger ($E_{\text{GW,post}}$) versus the instantaneous GW frequency ($f_{\text{GW,post}}$), averaged over the last 5 ms of the simulations.
We used the same EOSs as in Fig.~\ref{fig:f2vslambda} for NS (with the addition of some non-equal-mass models from~\citep{Rezzolla:2016nxn}) and NQS models. It is apparent that $E_{\text{GW,post}}$ is linearly correlated with $f_{\text{GW,post}}$ for all models: $E_{\text{GW,post}} = 4.308 \times 10^{-5}~(f_{\text{GW,post}} - 1687)$~($M_{\odot} c^2)$. The total GW energy emitted in the $f_2$-mode is $\sim 0.02-0.08 ~ M_{\odot} c^2$. The 1PT-NQS/QHC EOSs are clustered around the high/low end in the frequency-energy plane, respectively. The relation is not sensitive to the presence of quark degrees of freedom, via 1PT or crossover, nor to the binary mass ratio. 
Also note that the difference in energy emission in GWs between medium- and high-resolution simulations is approximately $5\%$ on average. Consequently, our conclusions remain unaffected by uncertainties arising from grid resolution.

\subsection{Relation between radiated energy and angular momentum after the merger}
Finally, we comment on the relation between the energy and the angular momentum radiated in GWs 
that was recently highlighted in~\cite{Ecker:2025puv}
(see also~\cite{Bernuzzi:2015opx, Chaurasia:2020ntk}). It is suggested that, at late times, different EOSs can be distinguished in such plots as they follow straight lines deviating from each other. 
While ~\cite{Ecker:2025puv} considers normalized energy and angular momentum emitted during the whole simulation, we propose using the angular momentum and energy emitted only after the merger (namely, subtracting the inspiral contribution) and without normalization, as shown in Fig.~\ref{fig:EvsJ-SupplementalMaterial}. 
This is because the normalization procedure with respect to the value at the time of merger (here defined, as usual, as the time of the peak amplitude of the GW strain) could be arbitrary. For example, if the same binary system is simulated for different numbers of orbits, the value of the angular momentum and energy radiated until merger would be different.
In practice, we suggest using the relation $E_{\text{GW}} - E^{\text{insp}}_{\text{GW}}$ vs. $J_{\text{GW}} - J^{\text{insp}}_{\text{GW}}$, where $E^{\text{insp}}_{\text{GW}}$~($J^{\text{insp}}_{\text{GW}}$) is the total energy~(angular momentum) carried away by GWs in the inspiral phase (See Appendix~\ref{Appendix-angmom-energy--details} for details).

\section{Summary.}
In addition to some hadronic NS mergers, in this Letter, we presented results from fully general-relativistic simulations of binary NQS mergers with EOSs compatible with current multi-messenger constraints and in which quark degrees of freedom are always present in isolated stars because of a 1PT or QHC~\citep{Kojo:2021wax}. 
We provided analyses that discriminate between many of the EOSs we studied. In particular, we showed that, in the $f_{2, \text{mean}}-\Lambda$ relation, QHC models stand out in general, and many of them lie significantly far from the best-fit curve for the hadronic models,
suggesting that the presence of a QHC could be identified if both inspiral and post-merger GW signals are detected.
More in detail, the following conclusions are evinced from our work: 
\begin{itemize} 

    \item Observationally confirming quark degrees of freedom in stable NSs usually requires signaling a transition from conventional hadronic stars to hybrid stars with quark cores. In other words, the scenarios where quarks are {\it always} or {\it never} present in NSs cannot be distinguished with observations of stars in equilibrium. We propose to call these models neutron-quark stars.

    \item  In the QHC scenario, one will not see $f_2$-mode GW emission for the most compact models because when densities in the core surpass those of the peak of the sound speed, the soft quark EOS plays a dominant role, and the merged object collapses in a short time, before any $f_2$-mode radiation is emitted.
    
    \item We construct the $\Lambda - f_2$ relation based on observationally favored hadronic EOSs. While current uncertainties remain significant, our method has the potential to distinguish hadronic vs. QHC EOSs in heavier stars (namely, except in our M1.25 models).

    \item The 1PT scenario deviates from the hadronic $M/R - \Lambda$ relation and exhibits a pronounced high-frequency shift ($\sim$ 300 Hz ) from its hadronic baseline (ChEFTex EOS) in the $f_2$ mode. However, it is close to hadronic models in the $\Lambda - f_2$ relation. More observable signatures are expected from the finite-temperature phase transitions.

   % \item Our discussion is based on future measurements of $f_2$-mode gravitational radiation. Likely, these will be carried out several years from now, and, by that time, NS observations through X-rays will have significantly reduced uncertainties in universal relations involving hadronic EOSs, thereby making deviations caused by QHC EOSs (Fig.~\ref{fig:f2vslambda}) even more pronounced.

\end{itemize}

Our analysis relies on prospective measurements of post-merger GWs of the $f_2$ mode. When the $f_2$ mode is clearly identified, the GW SNR will be sufficiently high to enable precise constraints on the tidal deformability from the inspiral phase~\citep{Flanagan1998PhRvD..57.4535F,Torres-Rivas2019PhRvD..99d4014T}. Furthermore, the universal relation involving hadronic EOSs in Figure~\ref{fig:f2vslambda} — along with its current uncertainty range — is derived from current observations. In the era of detectable post-merger GWs, tighter observational constraints on hadronic models are expected to substantially reduce the uncertainty associated with this universal relation. Consequently, the NQS scenario will become more easily distinguishable. However, integrating numerical simulations with multi-messenger observations, we demonstrated that the NQS scenario is in principle observable in certain cases, even given the current level of observational uncertainties.

Effects like those of strong magnetic fields or viscosity in the post-merger phase have been recently found to shift $f_2$ to higher values~\citep{PhysRevLett.134.121401,   PhysRevLett.134.071402}. This shift may be degenerate with the phase-transition scenario at finite temperature or may partially cancel the shift to lower frequencies seen in QHC models. We acknowledge the importance of quantitatively accounting for these potential effects and we will address these issues in future work. Currently, while cold EOSs can be studied statistically ({\it e.g.}, via Bayesian inference with multi-messenger data), research with numerical simulations on finite-temperature EOSs, magnetic fields, and transport properties is still limited to case studies.

In addition, we found a good linear correlation between the average instantaneous frequency and energy emitted in post-merger GWs. This could be used to estimate the emitted GW energy also in cases where only $f_2$ is measured accurately, rather than the whole post-merger strain.

Finally, we proposed a small modification to a recently proposed relation between the energy and angular momentum emitted in GWs. Our suggestion makes the relation less dependent on the initial conditions of the simulations and seems to highlight even further the differences due to EOSs.

Straightforward extensions to our work include studying the effects of varying the onset density of the 1PT and the binary mass ratio. Much more demanding, and interesting, would be extending the present analyses to NQS EOS at finite temperature. 

\section*{Acknowledgments}

 Our simulations were carried out on the Hokusai Bigwaterfall supercomputer in RIKEN and the XC50 system at the Center for Computational Astrophysics~(CfCA) of the National Astronomical Observatory of Japan~(NAOJ). Y.H. is thankful for the stimulating discussions at the 2024 TDLI Workshop on Dense Matter Equation of State and Frontiers in Neutron Star Physics. S.H. is supported by the Japan Society for the Promotion of Science~(JSPS) KAKENHI Grant No. 22F22750.  Y.H. is supported by the National Natural Science Foundation of China No. 12233011 and by the Postdoctoral Fellowship Program of CPSF under Grant Number GZC20241915. T.K. is supported by the JSPS KAKENHI Grant No. 23K03377 and No. 18H05407 and by the Graduate Program on Physics for the Universe (GPPU) at Tohoku University. H.S. is supported by the JSPS KAKENHI Grant No. JP23K20848 and JP24KF0090, and by the FY2023 RIKEN Incentive Research Project. H.S. and S.N. are supported by the Pioneering Program of RIKEN for Evolution of Matter in the Universe (r-EMU). K.T. is supported by the JSPS KAKENHI Grant No. 17K14305 and 23K03399. L.B. and S.N. are supported by the JSPS KAKENHI Grant No. JP25H00675. S.N. is supported by the JSPS KAKENHI Grant No. JP23K25874. S.N. is supported by the JST ASPIRE Program "RIKEN-Berkeley mathematical quantum science initiative".
 
\appendix
\section{Numerical Setup}
\label{Numerical Setup}
We produce the initial configurations for our simulations using the multi-domain pseudo-spectral open-source code \texttt{Lorene}~\citep{Gourgoulhon2001}. These are quasi-equilibrium irrotational~\citep{Cook-2000LRR3} equal-mass BNS configurations with an initial separation of 45 km, resulting in approximately 8 to 12 orbits before the merger, depending on the model. 

As in~\citep{HuangPRL2022}, we use the open-source high-resolution shock-capturing fully general-relativistic hydrodynamic code \texttt{WhiskyTHC}~\citep{10.1093/mnrasl/slt137, Radice:2013xpa}, which is based on the \texttt{Einstein Toolkit}~\citep{EinsteinToolkit:2023_05} framework.
Specifically, our computational approach for hydrodynamics involves the utilization of a finite-volume scheme featuring advanced 5th-order monotonicity-preserving reconstruction techniques~\cite{Suresh1997}. 
The Riemann problems within the scheme are handled with the Harten-Lax-van Leer-Einfeldt (HLLE) Riemann solver~\cite{harten1983upstream}. 
The evolution of spacetime is computed utilizing the Z4c formulation~\cite{Bona:2003fj, Bona:2004yp} through the \texttt{CTGamma} code, which implements ``1 + log" slicing for the lapse function and ``Gamma-driver" shift conditions~\cite{Alcubierre:2002kk}. The time integration of the hydrodynamics and Einstein equations is carried out with the method of lines, employing a third-order strong-stability-preserving Runge-Kutta scheme~\cite{Gottlieb2009}, chosen for its efficacy in maintaining stability. We set the Courant-Friedrichs-Lewy (CFL) factor to 0.075. This stringent choice is crucial when implementing flux reconstruction in local-characteristic variables, a technique integral to our adopted monotonicity-preserving scheme~\cite{Suresh1997}. 

For adaptive mesh refinement~(AMR), we use the \texttt{Carpet} code~\cite{ErikSchnetter_2004}, with seven refinement levels. 
Our outer boundary is set at 1477 km. 

We examined the dependence of $f_2$ on numerical resolution by comparing three different resolutions: Low (finest grid spacing $\Delta x \approx 369 \, \text{m}$), Medium ($\Delta x \approx 231 \, \text{m}$), and High ($\Delta x \approx 185 \, \text{m}$). In Table 1, models for which the resolution is not reported refer to our Medium resolution.

\section{Equations of State}
\label{Appendix-EOS-details}
{\it Nuclear equations of state.} We consider a nuclear EOS with leptons, the Togashi EOS~\citep{Togashi:2017mjp}, and the ChEFTex EOS based on Refs.~\citep{Lonardoni2020, Drischler2021a, Drischler2021b}.
The Togashi EOS is based on variational calculations with microscopic two- and three-body nuclear forces constrained by scattering experiments and spectroscopy of light nuclei. The EOS tables are given from the crust to very high-density domains with consistent use of nuclear forces.

The ChEFTex EOS is based on the ChEFT calculations for nuclear liquid from baryon number density $n_B = 0.5n_0 $ to $1.5n_0$ with supplemental treatments for $n_B < 0.5n_0$ and $n_B > 1.5n_0$. 
For $n_B < 0.5n_0$, the nuclear liquid picture is not trustable and we use a crust EOS. For $n_B > 1.5n_0$, ChEFT is not valid so EOS tables for $n_B >1.5n_0$ are based on extrapolation: we use a parametrization up to cubic polynomials, $\varepsilon = \sum_{n=0}^3 c_n (n_B/n_0)^n$, with the coefficients $c_n$'s determined by the fit to quantities provided in ChEFT in the range $[0.5, 1.5]n_0$.

Roughly speaking, the power of density, $N$, characterizes $N$-body short-range repulsion.
At very large density, the energy density scales as $\varepsilon \sim c_3 (n_B/n_0)^3 $ and the corresponding pressure is $P= n_B^2 \partial(\varepsilon/n_B)/\partial n_B \sim 2\varepsilon$. 
Thus, the sound speed asymptotically approaches $c_s^2 \sim 2~c^2$.
The dominance of three-body forces and the associated growth of $c_s^2$ are helpful to satisfy the $2M_\odot$ constraints, but this eventually violates the causality bound. 
In the ChEFTex EOS the causality violation occurs at $\simeq 6.4 n_0$, and, for the Togashi EOS, at $\sim 5.5n_0$. For both EOSs, such densities are not reached in our simulations.
Since nuclear EOSs predict a gentle stiffening with increasing density, the radius of higher-mass ($2.1M_\odot$) stars is slightly smaller than that of lower-mass ($1.4M_\odot$) stars: $R_{1.4} - R_{2.1} \simeq 1$ km.

{\it Hybrid equations of state with a first-order phase transition.} In our hybrid EOS with a 1PT, 1PT-NQS, we use the ChEFTex EOS up to $1.8n_0$ and an EOS with constant sound speed at densities past the transition.
In order to avoid significant violations of the observational constraints while keeping the 1PT nature visible (a 1PT at densities near the maximum allowed for NSs would also be consistent with current observations but would lead to prompt collapse after the merger and, thus, to no post-merger GW radiation), 
we choose $c_s^2=2/3~c^2$ in the quark matter part while demanding the 1PT to occur at $\simeq 1.8n_0$ and end at $\simeq 2.75n_0$.
In this setup, the stars in our simulations already contain quark matter in their core before the merger, and the stellar radius at $\sim 1.4M_\odot$ is smaller than the one for the ChEFTex EOS.
This is a novel choice for initial conditions for numerical-relativity simulations; to the best of our knowledge, in all previous general-relativistic numerical work on BNSs with 1PT EOSs, the transition occurred after the merger.
Since it is challenging to produce, with codes based on spectral methods, initial data (for simulations) that contain discontinuities, we resorted to smoothing out the discontinuities in $c_s^2$ (see Fig.~1 of the main text). 
To avoid any thermodynamic inconsistency, such smoothing was done at the level of $P(\mu_B)$, and we derived all the other thermodynamic quantities from the differentiation of $P(\mu_B)$ with respect to $\mu_B$.

In this work, the temperature dependence of the hadron–quark phase transition boundary was not explicitly incorporated into the phase transition model, as it could introduce additional uncertainties not constrained by NS observations. The onset density of the phase transition—defined as the critical baryon density at which quark-matter formation begins—is known to depend on temperature. This is evidenced by the negative slope of the phase boundary in the $T$–$\mu_B$ plane (see, e.g, Figure 1 in~\citep{Baym_2018}), which arises from the constant-pressure condition $dP = n_B d\mu_B + s dT = 0$ along first-order phase transition curves, yielding $d\mu_B/dT = -\Delta s_m / \Delta n_B$, where $\Delta n_B > 0$ (indicating that the density of quark matter exceeds that of hadronic matter) and $\Delta s_m > 0$ (reflecting the higher entropy in quark matter due to additional color and flavor degrees of freedom). The temperature dependence of the hadronic number density at the phase transition, $dn_h/dT$, is primarily governed by the term $(\partial n_h / \partial \mu_B)_T \cdot d\mu_B/dT$, which is negative under typical conditions, given that the contribution from $(\partial n_h / \partial T){\mu_B}$ is generally negligible at high chemical potentials or temperatures~\citep{Togashi:2017mjp,LATTIMER1991331}. This result implies that the phase transition occurs at lower densities as the temperature increases. Consequently, the softening of the EOS due to a finite-temperature PT may lead to more compact remnants in BNS mergers, potentially increasing the $f_2$-mode GW frequency of the post-merger object.

{\it Quark-hadron--crossover equations of state.} QHC EOSs~\citep{Kojo:2021wax} are produced by using different combinations of nuclear and quark-matter EOSs in the lower- and upper-density ranges.
We use a nuclear EOS for $n_B \le 1.5n_0$ and a quark-matter EOS for $n_B \ge 3.5n_0$; in between, the EOS is constructed by interpolating the nuclear and quark-matter EOSs. 
The interpolating functions are polynomials $P(\mu_B) = \sum_{n=0}^5 c_n \mu_B^n$ with the coefficients that are demanded to match the nuclear and quark-matter EOSs up to second-order derivatives. 
The quark EOS contains two variable parameters for the effective repulsion and the attractive correlations in diquark channels~\citep{Kojo:2021wax}.

The reader is referred to the main text to see the abbreviations we use for the different variants of QHC EOSs and 1PT EOS. The QHC EOSs used in this work are publicly available through the \texttt{CompOSE} database~\href{https://compose.obspm.fr/}{https://compose.obspm.fr/}.

\section{Gravitational-wave energy and angular momentum}
\label{Appendix-angmom-energy--details}

\begin{figure}
    \centering
    \includegraphics[width=0.23\linewidth]{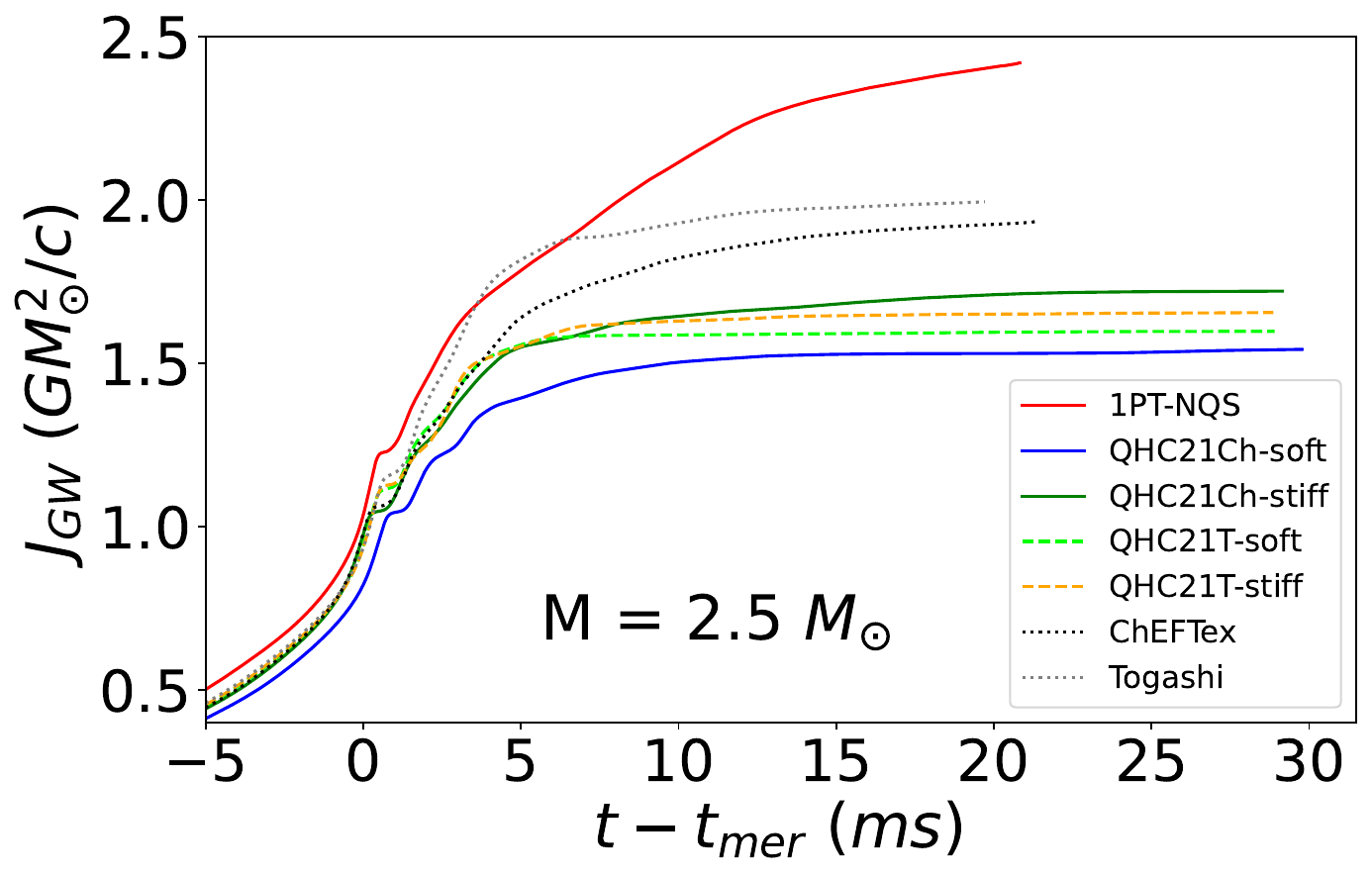}
    \includegraphics[width=0.23\linewidth]{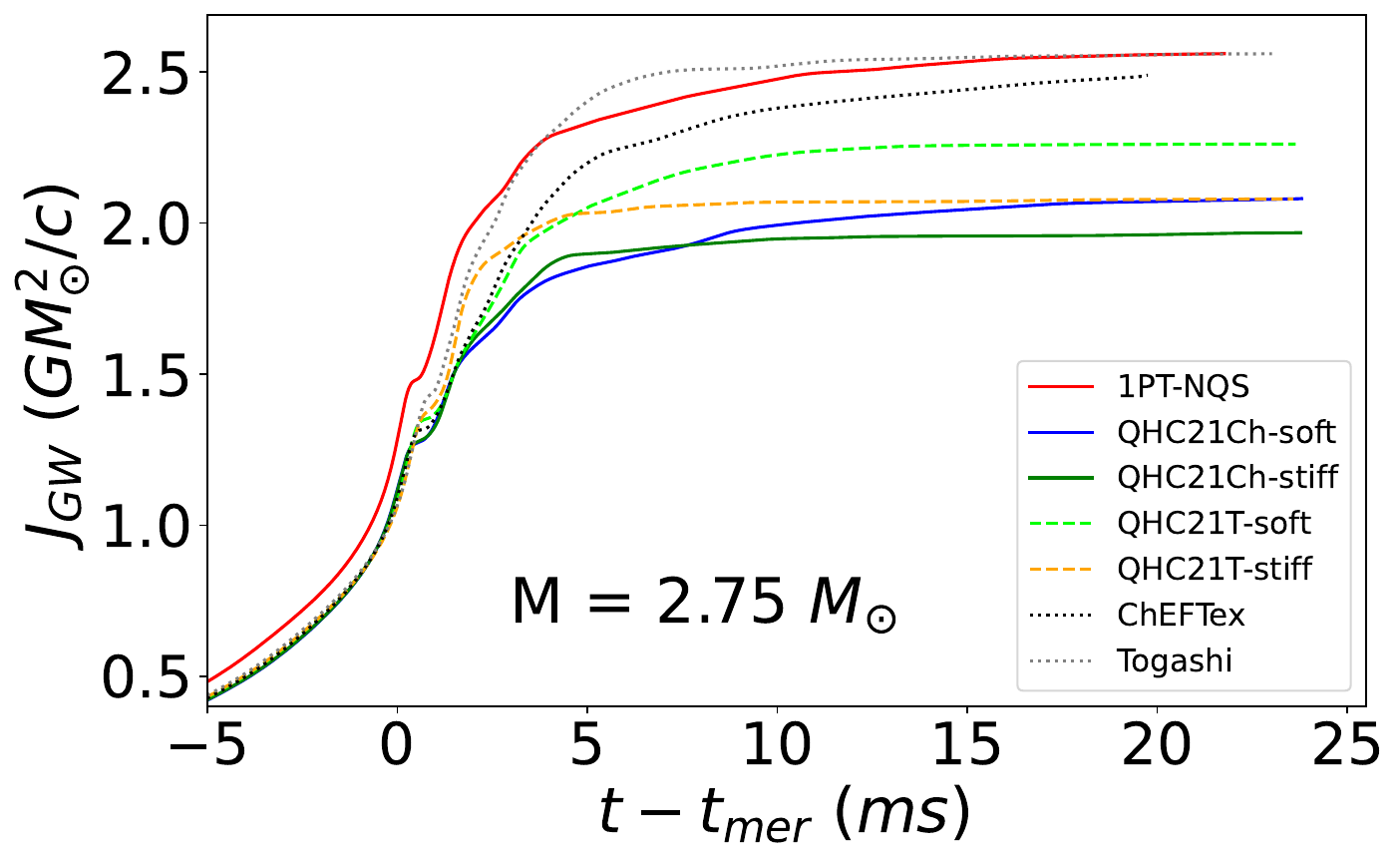}
    \includegraphics[width=0.23\linewidth]{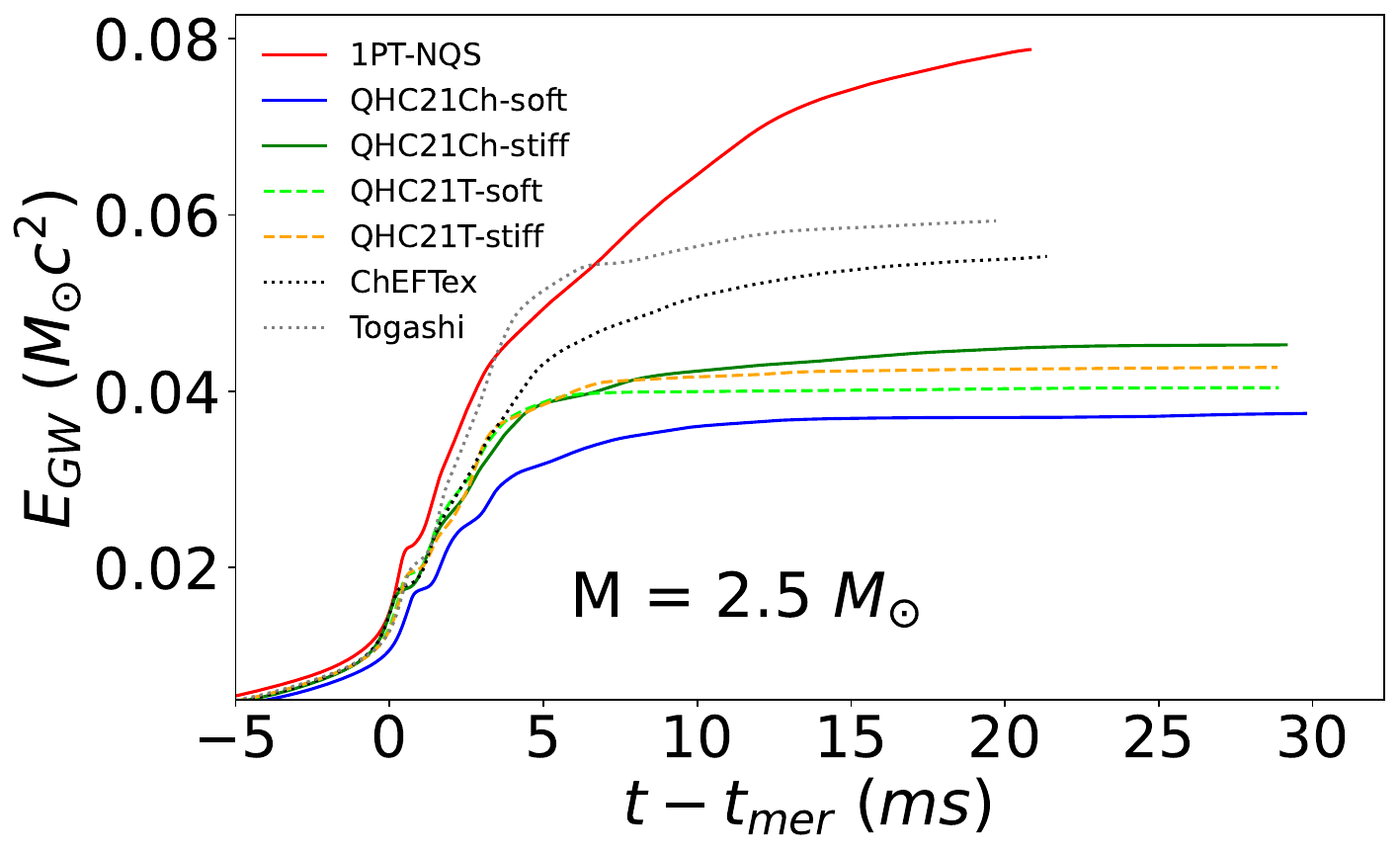}
    \includegraphics[width=0.23\linewidth]{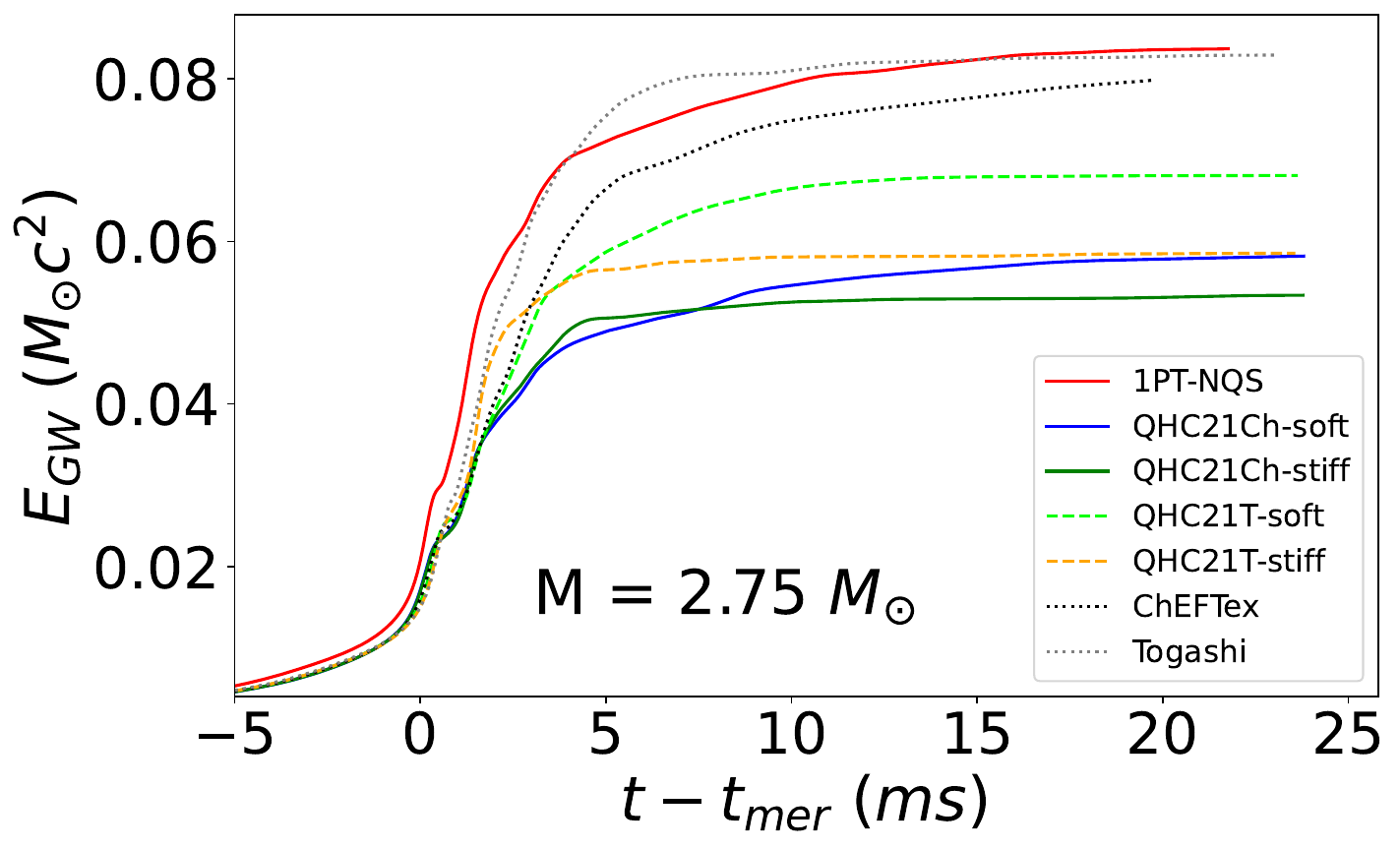}
    \caption{Radiated angular momentum and energy, as a function of time. $t=t_{\text{mer}}$ is the time of merger.}
    \label{fig:angmom-energy-unnorm}
\end{figure}

In this Appendix, we give further details on our considerations about energy and angular momentum radiated after the merger.
The total GW energy and angular momentum emitted after the merger do not depend on the lifetime of the remnant before collapse. This is because the amplitude of the dominant post-merger GW emission mode\footnote{The other radial-rotation coupling modes~\citep{Takami:2014zpa}, mentioned in the main text as $f_1$ and $f_3$, are active only in the first  $\approx 5$ ms and are less energetic.}, of frequency $f_2$, has similar evolution in all cases: it decreases steadily, and, as shown in Fig~\ref{fig:angmom-energy-unnorm}, GW emission becomes negligible $\approx 20$ ms (or less) after the merger, even if the remnant survives longer. These observations both led us to establish and justify the universal relationship shown in Figure~\ref{fig:E_fins} and led ~\cite{Ecker:2025puv} to propose their method to distinguish EOSs during the late post-merger phase, since in plots of radiated energy vs. radiated angular momentum, different EOSs can be distinguished as they follow straight lines deviating from each other. 

\begin{figure}
   \centering
   \hspace{0.3cm}
   \includegraphics[width=0.44\linewidth]{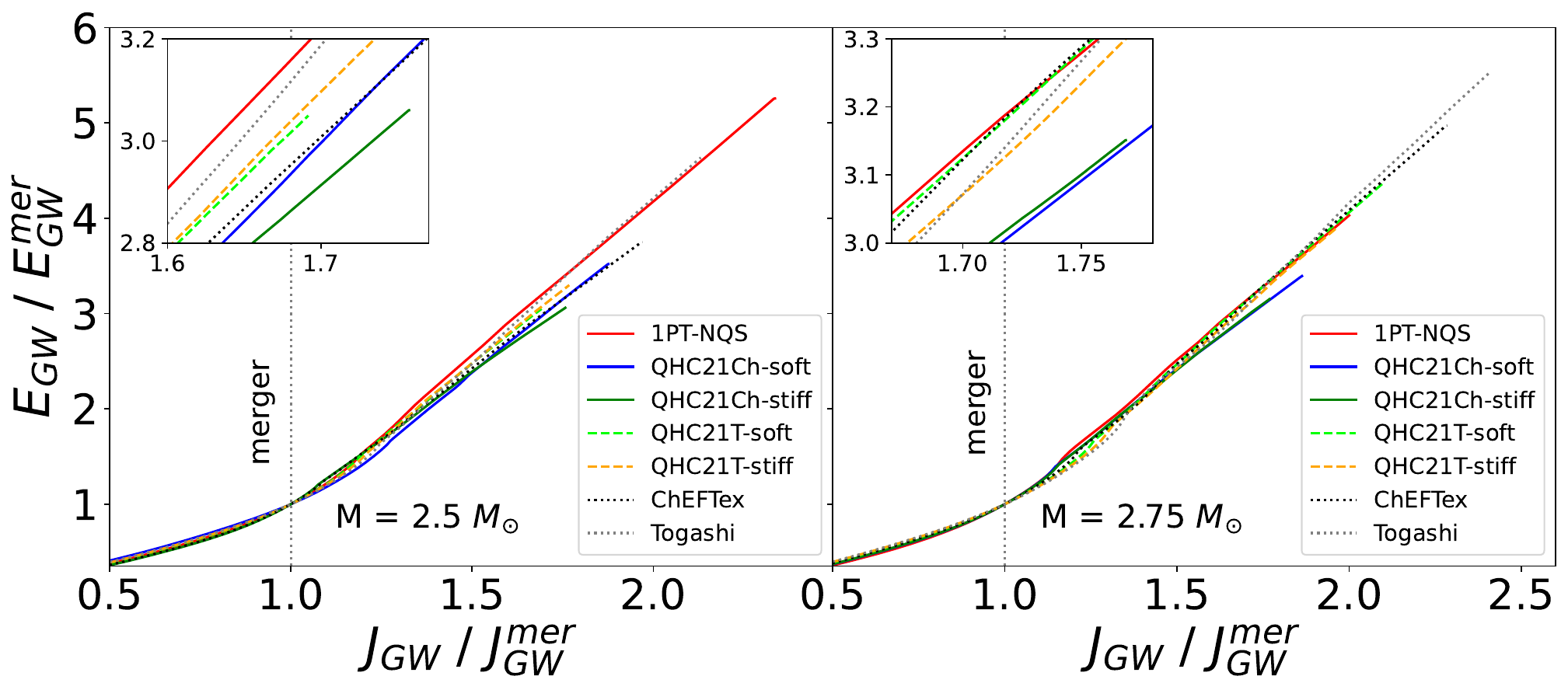}
   \includegraphics[width=0.46\linewidth]{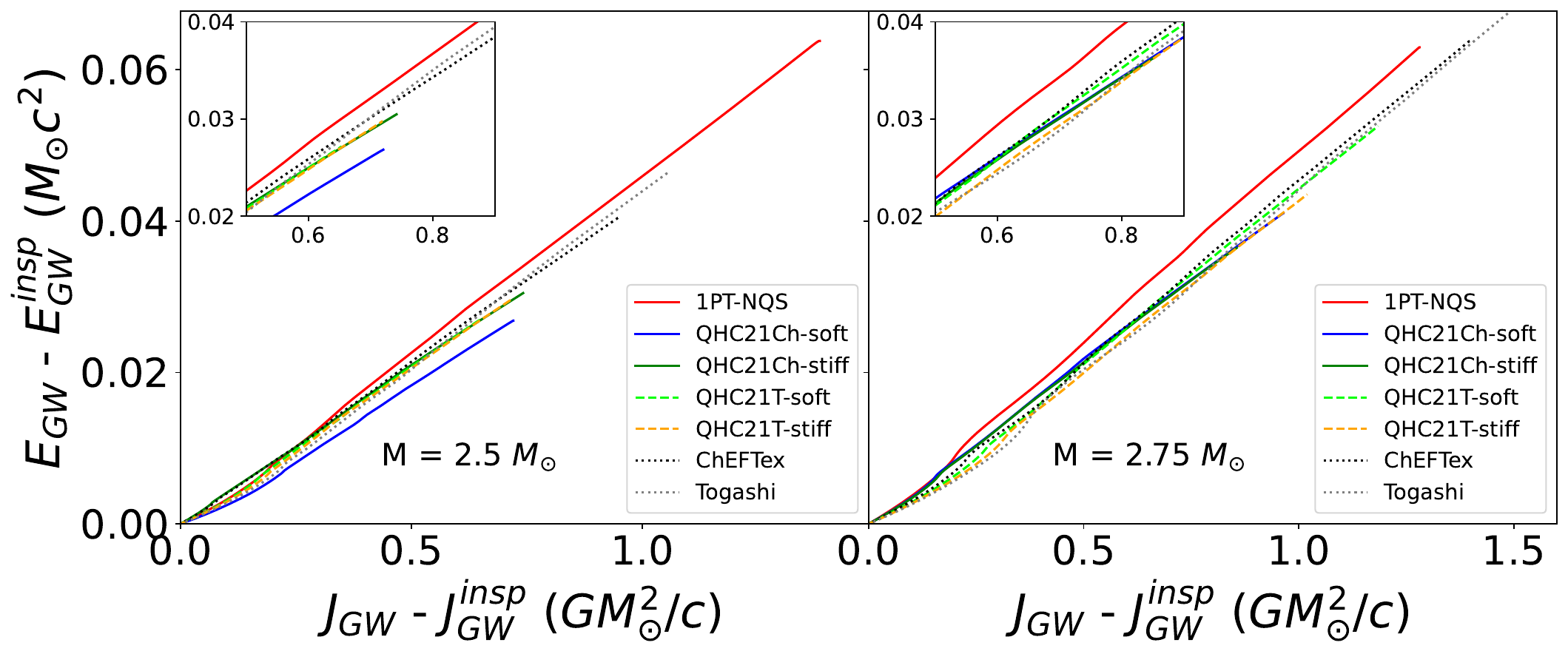}
   \caption{Radiated energy as a function of radiated angular momentum. The two panels on the left show the relation between quantities normalized to their values at the time of merger. The two panels on the right show non-normalized quantities calculated excluding the inspiral contribution.}
   \label{fig:EvsJ-SupplementalMaterial}
\end{figure}

In Fig.~\ref{fig:EvsJ-SupplementalMaterial}, we show such relations, in different forms. The left panels show the quantities computed as in~\cite{Ecker:2025puv}, namely, normalized to their values at the time of merger. The right panels show our suggestion of considering non-normalized post-merger-only quantities. 
In the low-mass case with normalized quantities, the 1PT-NQS EOS and some of the QHC models can be somewhat distinguished. However, for the intermediate-mass case, the deviations are much smaller. The reason for this could be that we stopped our intermediate-mass simulations earlier. Longer post-merger simulations may be required to better reproduce the results of previous studies~\citep{Ecker:2025puv}. In any case, we can see that in the relation that considers non-normalized post-merger-only quantities, the curve for each model deviates even more from those of the other models, especially in the intermediate-mass case.

\section{Lifetime dependence on grid resolution and thermal index}
\label{Appendix-lifetime--details}

The lifetime of the merger remnant, defined as the time between the merger and the subsequent collapse into a black hole, holds significant scientific interest due to its direct connection to the physics of the post-merger, including the thermal behavior of the EOS, magnetic field evolution, and neutrino transport. However, accurately estimating this timescale in numerical simulations remains challenging also because of sensitivity to numerical setups like resolution. Despite this, \cite{Harada-Cannon-etal_2024PhRvD.110l3005H} found that it is reasonable to expect to be able to identify the correct transition scenario with third-generation detectors or specialized detectors with high sensitivity at high frequencies.
We were inspired by the results discussed in~\cite{Fujimoto:2022xhv} to study the lifetime of the merged object,  with our setup and EOSs. In~\cite{Fujimoto:2022xhv}, a {\it weak} dependence of the lifetime on $\Gamma_{\mathrm{th}}$ and grid resolution was reported.
That being a qualitative study of this point, we were motivated to test the assertions quantitatively.
As discussed earlier, the effective temperature treatment performed by adding a {\it hot} component to the EOS through an ideal-fluid EOS with index $\Gamma_{\mathrm{th}}$ increases pressure support in the merged object.
Therefore, a change in the value of the $\Gamma_{\mathrm{th}}$ index directly impacts the lifetime.

\begin{figure}[h]
    \centering
    \includegraphics[width=0.45\linewidth]{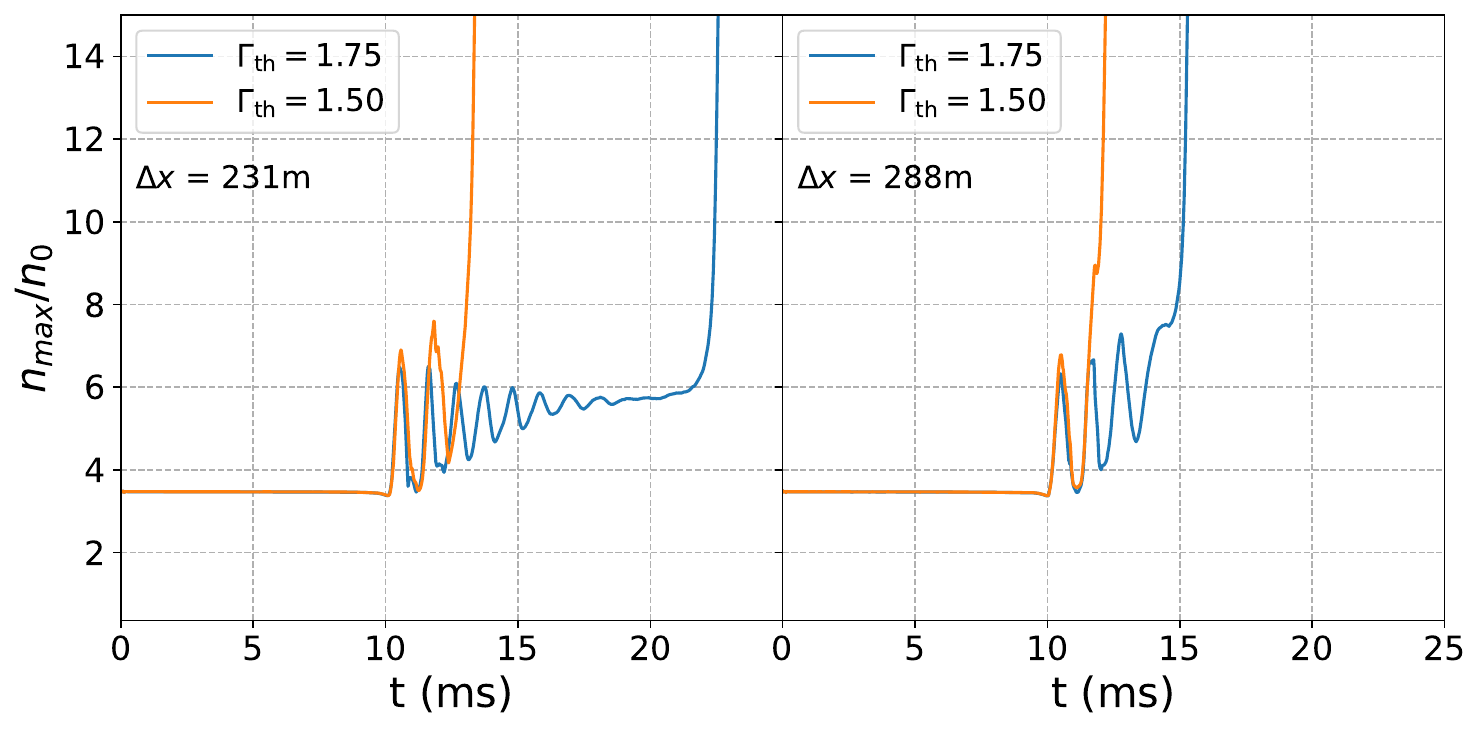}
    \includegraphics[width=0.45\linewidth]{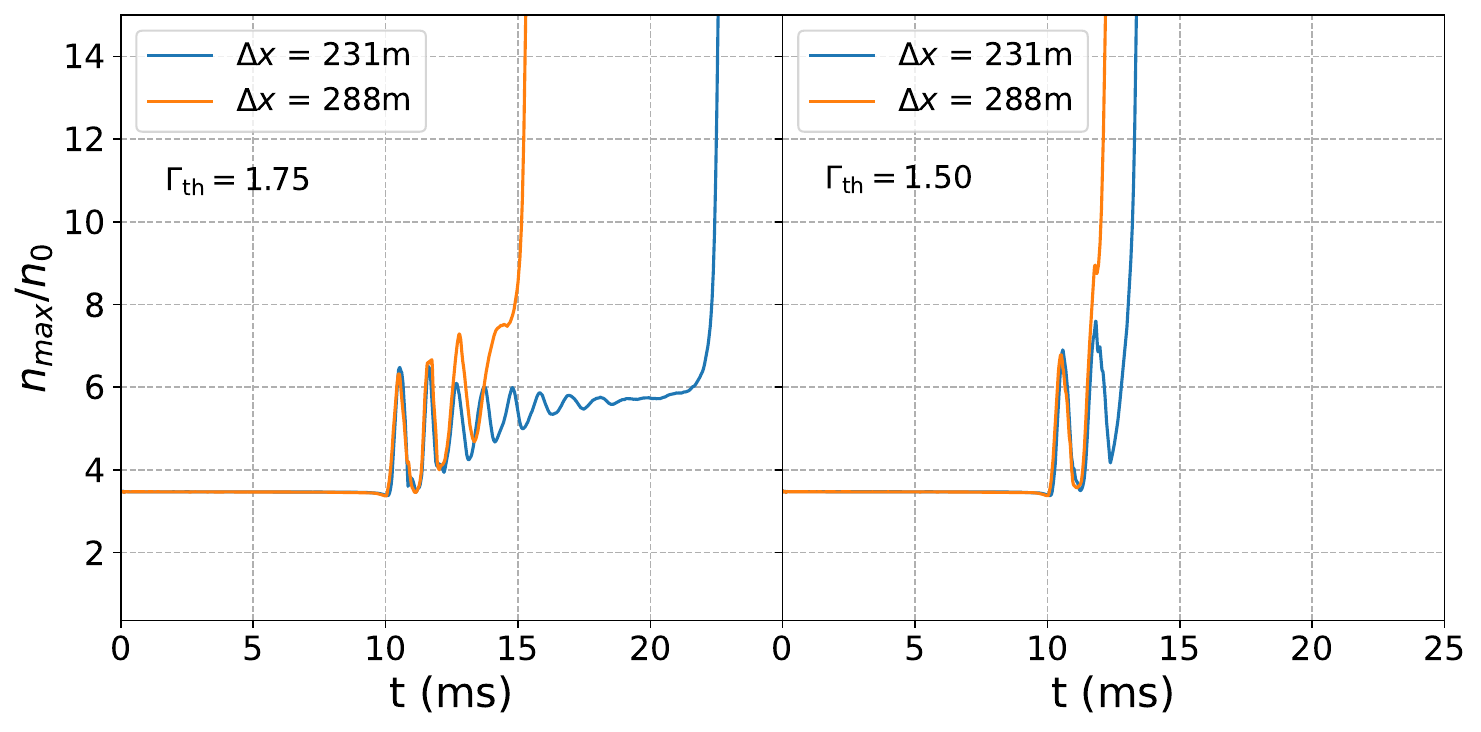}
    \caption{Evolution of the maximum of the number density~($n_\text{max}$) in units of nuclear saturation density for the QHC21Ch-stiff EOS model with mass 3.1$M_{\odot}$ and different combinations of $\Gamma_{\mathrm{th}}$ and grid spacing of the finest refinement level, $\Delta x$.}
    \label{fig:lifetime}
\end{figure}

Furthermore, it is interesting to quantitatively investigate the influence of grid resolution.
To have a similar setup to that of~\cite{Fujimoto:2022xhv}, we investigated the cases of $\Gamma_{\mathrm{th}}=1.5$ and $\Gamma_{\mathrm{th}}=1.75$. 
We performed simulations of a binary system of mass 3.1$M_{\odot}$ with the QHC21Ch-stiff EOS, and at two resolutions: finest grid spacing of 231 m and 288 m. 
This mass was selected to have a timescale for the lifetime comparable to that of the aforementioned study.
We let the evolution continue until an apparent horizon \citep{Thornburg2007,Thornburg2004:multipatch-BH-excision} was found.
The evolution of the maximum number density for different configurations is shown in Fig.~\ref{fig:lifetime}. From the two panels on the left, one can compare the lifetime for different values of $\Gamma_{\mathrm{th}}$ and for a given grid resolution.
The leftmost panel points out a notable 41\% difference in the lifetime between the cases of different $\Gamma_{\mathrm{th}}$ for the higher resolution, whereas this difference decreases to approximately 20\% for the lower resolution.
In the two panels on the right, the lifetime for a choice of $\Gamma_{\mathrm{th}}$ with different finest-grid spacings is presented. The change in lifetime is about 32\% for different resolutions for $\Gamma_{\mathrm{th}}=1.75$, while it is 9\% for $\Gamma_{\mathrm{th}}=1.5$.

\begin{figure}[h]
    \centering
    \includegraphics[width=0.98\linewidth]{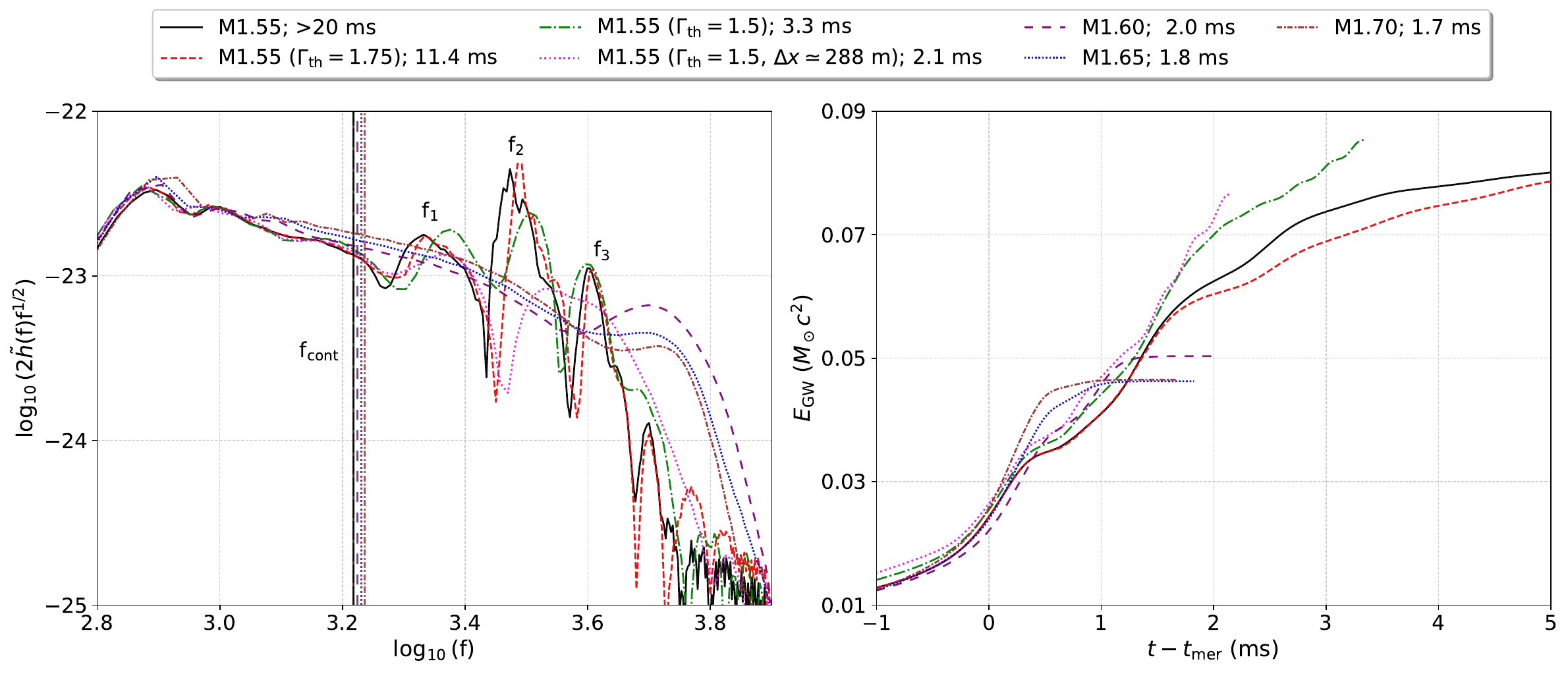}
\caption{
{\it Left panel:} Power spectral densities for the QHC21Ch-stiff model. 
Here,  $\tilde{h}(f) \equiv \sqrt{\frac{\left| \tilde{h}_{+}(f) \right|^2 + \left| \tilde{h}_{\times}(f) \right|^2}{2}}$, where $\tilde{h}_{+}$ and $\tilde{h}_{\times}$ are the Fourier transforms of the two polarization modes ($h_{+}$ and $h_{\times}$), computed as we did in~\cite{HuangPRL2022}. Vertical lines indicate the contact frequency. 
The labels in the legend consist, for each QHC21Ch-stiff model, in the rest mass of a star in the binary, followed by the $\Gamma_{\text{th}}$ coefficient if different from 2, followed by the finest grid spacing if different from the standard 231 m, followed by difference between the time between merger ($f_{\text{cont}}$) and collapse.
{\it Right panel:} Energy radiated in the $l=m=2$ GW mode as a function of time; $t_{\text{mer}}$ is the time of merger.
}
    \label{fig:psd}
\end{figure}

\section{No-$f_2$ mass threshold}
\label{no-f2 mass threshold}
In BNS merger simulations, it is customary to define {\it prompt collapse} as collapse occurring during the first increase of the maximum density in the system being simulated. Thus, prompt collapse cases are expected to release zero GW energy in the post-merger phase.
In some cases, there may be post-merger GW radiation that, however, does not show $f_2$-mode emission (see the left and right panels of Fig.~\ref{fig:psd}): Over a very short lifetime, GWs may be released in a continuous spectrum without post-merger peaks. 

Therefore, we define the {\it no-$f_2$ mass threshold} as the lowest mass of one star in the binary (only for equal-mass cases here) for which no $f_2$-mode GW emission is identified. Since $f_2$-mode emission is identified by the presence of a peak in the spectrum, the criterion for the no-$f_2$ mass threshold is chosen to be $\text{PSD}(f) < \text{PSD}(f_\text{cont})$ for all $f > f_\text{cont}$, where $f_\text{cont}$ is the contact frequency. The contact frequency is the orbital frequency at binary contact (when the relative distance between the centers of the stars is equal to the sum of the radii of the two stars). There is a useful analytical approximation for it $f_{\rm cont} = C^{3/2}/(2\pi M)$, where $C$ represents stellar compactness~\citep{Damour-contact-frequency}.
As seen in Fig.~\ref{fig:psd}, $f_\text{cont}$ identifies sufficiently well the start of the merger and is always lower than $f_2$. We chose $f_\text{cont}$ also because it is straightforward to estimate through its analytical approximation $f_\text{cont} = C^{3/2}/(2\pi M)$ (where C is the stellar compactness)~\citep{Damour-contact-frequency}; no fitting is required. 

As an additional comment, note that the mass threshold for prompt collapse is higher than the no-$f_2$ mass threshold.
Fig.~\ref{fig:psd} displays QHC21Ch-stiff configurations differing in mass, thermal index, or resolution. 
We can identify models that do not collapse promptly but that do not show $f_2$ in their spectrum. These are models with short lifetimes ($\sim 2$ ms). 

The simulation of the QHC21Ch-stiff configuration with stellar masses of $1.7 M_{\odot}$ shows no $f_2$-mode GW emission but does not collapse promptly. This points to a possible difference of $\sim 0.3 M_{\odot}$ in the total binary mass between the prompt-collapse mass threshold and that for the absence of $f_2$-mode GWs, with the no-$f_2$ mass threshold being at $\approx 1.60 M_{\odot}$.

\bibliography{testref}
\bibliographystyle{aasjournal}

\end{document}